\documentclass[12pt]{article}
\usepackage[top=2.5cm, bottom=2.5cm, left=2.5cm, right=2.5cm]{geometry}
\usepackage{graphicx}
\usepackage{booktabs}
\usepackage{amsmath,amsthm,amsfonts,amssymb,amscd}
\usepackage{titlesec}
\usepackage{enumitem}
\usepackage{mathtools}
\usepackage{cancel}
\usepackage{caption}
\usepackage{float}
\usepackage{array}
\usepackage[round]{natbib}
\usepackage{xcolor}
\usepackage{xurl}
\usepackage[colorlinks=true,allcolors=blue]{hyperref}

\usepackage[section]{placeins} 

\usepackage{afterpage}
\usepackage{authblk}

\title{Sea Surface Roughness Dependence on Ocean Wave Parameters through Large Eddy Simulation with Local Subfilter Wave Drag}
\date{\vspace{-5ex}}

\author[1*]{Hana Hata Williams}
\author[2]{Aditya K. Aiyer}
\author[1,3]{Luc Deike}
\author[1]{Michael E. Mueller}
\affil[1]{Department of Mechanical and Aerospace Engineering, Princeton University, Princeton, New Jersey, USA}
\affil[2]{Department of Mechanical Engineering and Mechanics, Lehigh University, Bethlehem, Pennsylvania, USA}
\affil[3]{High Meadows Environmental Institute, Princeton University, Princeton, New Jersey, USA}
\affil[*]{Correspondence: hhwilliams@princeton.edu}

\begin{document}
\maketitle

\begin{abstract}
Characterizing the Marine Atmospheric Boundary Layer (MABL) requires understanding the coupling between ocean waves and the turbulent atmospheric boundary layer above them. This coupling controls momentum exchange between the atmosphere and the ocean; it is of practical importance in the global climate, flow of ocean currents, ocean engineering, and offshore wind energy. Computational study of the MABL is complex because it must resolve the coupled physics of waves and turbulence over a wide range of spatial and temporal scales. This study expands on approaches for representing dynamic, local waves in Large Eddy Simulations (LES) of the MABL by developing a subfilter wave drag model to be local and scale-invariant. It explores the effects of different wave parameters (significant wave height and peak frequency of the wave energy spectrum) on the resulting momentum flux beyond monotonic relationships between surface stress through friction velocity $u_\ast$ and wind velocity above the surface $U_{10}$. Results are compared to field data and in a discussion on how representation of the MABL and associated momentum flux need to account for both wind and wave effects.
\end{abstract}


\subsection*{Significance Statement}
This study presents a new model for the drag imparted by small waves within an existing computational framework for representing the influence that ocean waves have in simulations of the wind above them. The model accounts for the fact that the impact of waves changes in both space and time and cannot be defined simply for an entire ocean or computational domain. The study then leverages the ability to change wind and wave parameters independently in simulations to isolate wave influences on the shape of the wind profile above the sea. Learning more about the drag at the air-sea interface has implications in large-scale ocean, atmosphere, and Earth system modeling as well as power prediction for offshore wind farms.

%

\section{Introduction}
Interactions between the Marine Atmospheric Boundary Layer (MABL) and ocean waves dictate the exchange of momentum, heat, and gases between the ocean and atmosphere, playing a key role in global climate. The transfer of momentum between air and sea is a key piece of the puzzle: not only is developing methods for accurately representing ocean waves in simulations of the MABL useful for many ocean engineering applications such as predicting power output of offshore wind farms \citep{veers_grand_2023}, but the transfer of momentum also contributes to the growth and breaking of waves, which in turn influence mass transfer at the surface \citep{deike_mass_2022}. Many theoretical, computational, and experimental studies have been devoted to this problem. 

Previous studies to clarify the relationship between the shape of the waves, the drag imposed by the waves on the wind, the wind speed, and the shape of the MABL have largely centered the assumption 
\citep{ayet_dynamical_2022} that the MABL takes a logarithmic profile \citep{pope_turbulent_2015}: 

\begin{equation}
    \label{eq:loglaw}
    U(z) = \frac{u_\ast}{\kappa}\log\left(\frac{z}{z_0}\right) \text{ ,}
\end{equation}
where $\kappa = 0.4$ is the von Karman constant and $z_0$ is the roughness height. The friction velocity $u_\ast$ is defined to relate to the surface stress $\tau_{tot}=\rho_a u_\ast^2$, 
where $\tau_{tot}$ is the total air-sea momentum flux. The logarithmic relationship is expected to be valid for a neutrally-stratified, quasi-stationary, homogeneous flow above the wave boundary layer \citep{hara_wind_2004} and may not be valid in conditions where waves heavily modify the flow \citep{buckley_surface_2020} or in strong swell conditions \citep{sullivan_large-eddy_2008}. 
One can additionally introduce a drag coefficient $C_D$:
\begin{equation}
    \label{eq:drag}
    \tau_{tot} = \rho_a u_\ast^2 = C_D \rho_a U_{10}^2 \text{ ,}
\end{equation}
where $\rho_a$ is the density of air and $U_{10}$ is the velocity 10 m above the sea surface. 

\begin{figure}[htbp]
  \centering
    \includegraphics[width=0.7\textwidth]{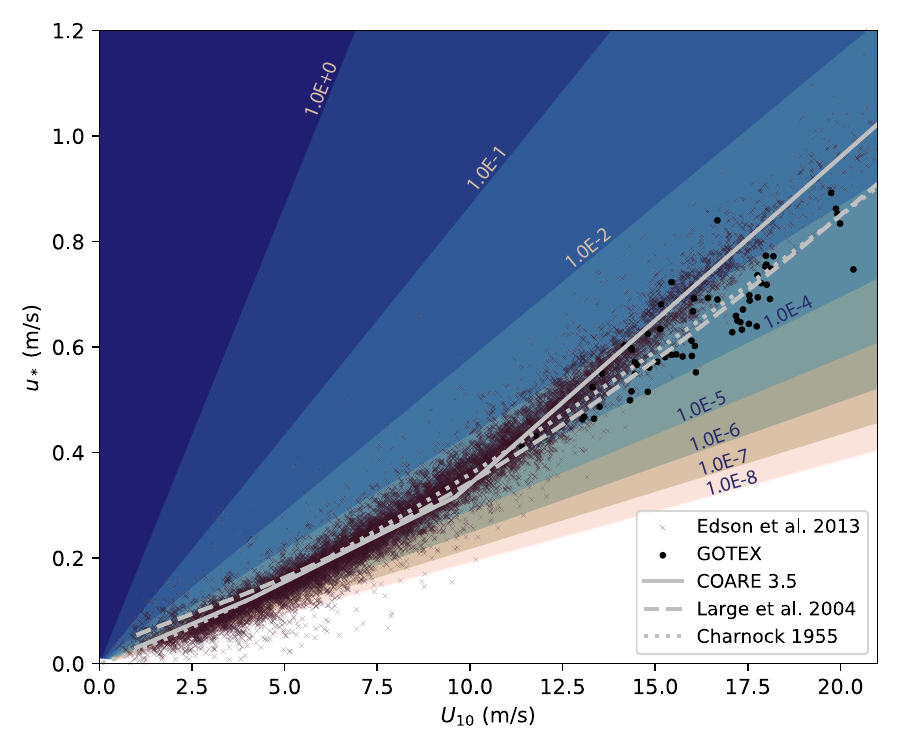}
  \caption{Contours of constant roughness $z_0$ on a plot of friction velocity $u_\ast$ as a function of the velocity 10 m above the sea surface $U_{10}$, as defined by the log law. The magnitude of $z_0$ (in m) is annotated above and to the left of each contour. The solid gray line indicates where MABLs would fall according to the COARE 3.5 fit \citep{edson_exchange_2013}. The dashed gray line does the same for the NCAR momentum flux formulation \citep{large_diurnal_2004}. The dotted gray line indicates MABLs according to the Charnock relation, $z_0 = \frac{\alpha_{ch}u_\ast^2}{g}$ with $\alpha_{ch}$ = 0.011 for the open ocean \citep{Charnock1955}, and the log law. Data marked by a red `$\times$' are from the aggregated data in \citet{edson_exchange_2013}, and data marked by black circles are from the GOTEX fetch-limited campaign \citep{romero_airborne_2010}.}
  \label{fig:z0contours}
\end{figure}

Under the assumptions of equations~\eqref{eq:loglaw} and~\eqref{eq:drag}, the questions of momentum flux formulation---how to define the drag coefficient $C_D$, wind friction velocity $u_\ast$, and roughness $z_0$---are all interconnected \citep{ayet_dynamical_2022}. The wind speed at 10 m above the sea surface $U_{10}$ is conventionally used in discussions of momentum transfer at the air-sea interface because it is assumed to be both above immediate wave effects and in the logarithmic region of the boundary layer. It is additionally a practical choice because it is close to the height at which sensors on ships or field platforms are placed. Figure \ref{fig:z0contours} shows a plot of $u_\ast$ as a function of $U_{10}$, colored by contours of constant $z_0$, to visualize the relationship defined by the log law. Many previous studies have attempted to constrain these relationships theoretically or empirically. The Charnock relation, for example, defines the roughness height based on friction velocity: $z_0 = \frac{\alpha_{ch}u_\ast^2}{g}$, where $g$ is the gravitational acceleration and $\alpha_{ch}$ is a non-dimensional constant \citep{Charnock1955}. This relationship assumes that the effect of waves on the momentum flux can be captured by a roughness that scales with the square of the wind friction velocity, or directly with the surface stress, a result derived from constructing a non-dimensional group with $u_\ast$, $g$, and $z_0$. To account for difference between wave conditions, the Charnock parameter can be defined differently based on coastal waters and open ocean \citep{Charnock1955}.
Some parameterizations of drag at the air-sea interface that vary linearly with the wind speed have been proposed, either through $C_D$ \citep{garratt_review_1977} or $u_\ast$ \citep{andreas_new_2012}. \citet{hsu_dynamic_1974} and \citet{donelan_dependence_1993} proposed that $z_0$ is a function of the significant wave height $H_s$ and the wave age $c_p/u_\ast$, but datasets did not produce consensus on constants, and possible spurious correlation from wave age being dominated by a $u_\ast$ tied directly to $z_0$ was raised \citep{DrennanTaylorYelland2005, andreas_new_2012}. \citet{taylor_dependence_2001} proposed a parameterization based on the wave steepness $H_sk_p$, but the parameterization was not shown to perform better than those based on the wave age \citep{DrennanTaylorYelland2005}. 
More recently, \citet{edson_exchange_2013} presented the COARE 3.5 parameterization, which is based on extensive field measurements of the wind speed at 10 m and the wind friction velocity and defines a wind-dependent $u_\ast(U_{10})$ in a piece-wise fashion. Another widely used formulation from the National Center for Atmospheric Research (NCAR) defines $u_\ast(U_{10}) = U_{10} \left(\frac{2.7\times10^{-3}}{U_{10}} + 7.64\times10^{-5}U_{10} + 1.42\times10^{-4}\right)^{1/2}$ \citep{large_diurnal_2004}. The Charnock, NCAR, and COARE 3.5 parameterizations are shown in Figure \ref{fig:z0contours} in comparison to field data from \citet{edson_exchange_2013}. Surface roughness parameterizations that take into account misalignment between wind and waves or mixed seas have also been proposed \citep{Porchetta2019, sauvage_improving_2023}; neither these nor other parameterizations including wave information have been adopted widely for large-scale modeling.

There has also been development of physically-based models that compute the drag by integrating the wave-induced stress over the wave spectrum \citep{janssen_quasi-linear_1991, makin_drag_1995, kudryavtsev_impact_2001}. These models treat the wave field as a superposition of linear, small-slope components and compute the form drag from the correlation between surface pressure and surface slope. For fully developed seas, the models recover the Charnock relation, and the predicted dependence of drag on wave age is sensitive to the assumed wave-age dependence of the high-wavenumber spectrum, which carries most of the wave-induced stress \citep{makin_drag_1995}. Formulations that incorporate increased complexity based on machine learning approaches \citep{wu_data-driven_2026} 
or based on a sea surface elevation map and corresponding wind behavior \citep{ayala_surface_2026} have recently been proposed as well. These approaches typically characterize a large surface (for example, a cell in a mesoscale model) through a single roughness parameter that averages out wave phase dynamics.

Field data for MABLs present significant scatter around fitted relationships that can be related to different sea states at any given wind speed, leading to variations in the momentum flux that point to additional dynamics that are not accounted for by the literature: see, for example, discussion in the review from \citet{ayet_dynamical_2022}. Most field data are taken on fixed, floating, or airborne platforms that, even with sophisticated instrumentation, have difficulty measuring the full vertical extent of the boundary layer at high resolution particularly close to the surface and in conjunction with wave conditions, so gathering consistent, accurate data at full scale in a variety of conditions is challenging. Isolating wind and wave conditions in the field is additionally complex; some field campaigns investigate specific configurations like the fetch-limited study by \citet{romero_airborne_2010}, while others rely on analysis through aggregation \citep{mahrt_sea_1996, hristov_dynamical_2003, edson_coupled_2007,edson_exchange_2013}.

Given challenges in performing field observations, simulations are an important tool for investigating the MABL but require an accurate representation of the momentum flux. A wide variety of approaches have been considered. A simple way to represent waves in Large Eddy Simulation (LES) is through the definition of a constant static roughness parameter $z_0$, such as that defined by the Charnock relation, and implementing a surface stress accordingly \citep{Moeng1984}. This method, which can be referred to as a wave phase-averaged approach because it averages out phase-dynamics of the waves through the single constant roughness, is frequently used in studies at the scale of an offshore wind farm because it does not require higher resolution for the waves \citep{Deskos2021}. This simple approach, however, does not adequately capture wave dynamics (see, e.g., Figure \ref{fig:z0contours}).

Separately there has, in the last couple decades, been much work to include the impacts of the waves in a more accurate, dynamic way by resolving the boundary between wind and wave through detailed simulations with a moving boundary \citep{sullivan_simulation_2000, yang_simulation_2011, yang_simulation_2011-1}; LES with a stationary rough surface \citep{anderson_dynamic_2011}; and LES with a moving boundary for monochromatic \citep{sullivan_large-eddy_2008} and a spectrum of waves \citep{yang_dynamic_2013, Sullivan2014, sullivan_turbulent_2018}, usually with some additionally prescribed subfilter roughness for any unresolved waves. 

Another large body of LES studies has investigated detailed statistics of the wave induced flow in a variety of configurations (misalignment, varying wave age, stratification). These studies are focused near the sea surface and have high vertical and horizontal resolution \citep{kihara_relationship_2007, suzuki_impact_2013, hara_wave_2015, wang_surface_2020, cao_simulation-based_2020, zhang_toward_2023,hao_windwave_2019, yang_effect_2022, husain_wind_2022, wu_revisiting_2022, zhu_moving-wave_2023, li_large-eddy_2025, scapin_momentum_2025}.

Complete understanding of the nature of drag between ocean waves and the full MABL remains elusive. One challenge in going from small-scale (lab or high-fidelity LES) to full-scale configurations is the large range of scales spanned in the ocean: longest wavelengths are up to hundreds of meters and MABL heights vary from a few hundred to two thousand meters. Many previously mentioned LES studies are at lab scales with vertical domain heights of one wavelength or a few hundred meters above the sea surface, depending on the chosen parameters, and do not capture the full shape of the MABL above the waves due to the computational expense of extending the domain. Many also validate either against lab experiments, where the wind and wave parameters can be controlled, or more resolved simulations. While phase-resolving simulations are more accurate than their phase-averaged counterparts and are frequently used as validation for fully wall-modeled simulations, their computational expense makes it difficult to scale up to a full-size MABL. 

Examining configurations that vary spatially or are not in equilibrium with the waves, as real oceans frequently are \citep{sullivan_large-eddy_2008}, requires tools that understand local dynamics. More recent studies have proposed wall models that incorporate the phase dynamics of a propagating wave spectrum without having to significantly increase the LES resolution, allowing for larger-scale study.
\citet{ayala_moving_2024} present the moving surface drag (MOSD) model, which computes surface drag based on local surface geometry of horizontally-resolved moving waves implemented in a flat-bottomed LES. The wave drag model developed by \citet{aiyer_dynamic_2024} compute resolved contributions to wave drag through the superposition of individual wave mode drag contributions. It comprises a fully wall-modeled representation of an ocean wave spectrum while preserving wave phase dynamics through the definition of a spatially- and temporally-varying drag force that is imposed on the bottom-most cell of the LES grid. It defines a locally-varying drag from resolved waves---those whose wavelengths are larger than the LES grid---and a subfilter drag for the unresolved wave modes based on 
an equilibrium log-law model and an equivalent subfilter wave roughness $z_{0,\Delta}$ \citep{bou-zeid_scale-dependent_2005, Piomelli2008}. Many previous LES with waves, particularly those with phase-resolved approaches where wave contours are resolved by the grid, have defined the subfilter roughness as a constant, universal value, frequently equal to $z_0 = 0.0002$ m \citep{sullivan_large-eddy_2008, Sullivan2014, hara_wave_2015, jiang_impact_2016, hao_windwave_2019, husain_wind_2022, zhu_moving-wave_2023, li_large-eddy_2025}. \citet{anderson_dynamic_2011} define the globally-averaged subfilter roughness for stationary rough surfaces as a function of a dynamic parameter $\alpha_w$ and the r.m.s.\ of the surface height fluctuation $\sigma_\eta^\Delta$ such that $z_0 = \alpha_w \sigma_\eta^\Delta$. \citet{yang_dynamic_2013} expand upon this model by exploring constructions for $\sigma_\eta^\Delta$ based on geometry, a steepness-dependent Charnock parameter, wave kinematics (accounting for wave propagation speed), and a combination of wave kinematics and steepness. The models for $\sigma_\eta^\Delta$ superpose contributions to the subfilter $z_0$ from individual sinusoidal wave modes; the authors show that among tested models, the wave kinematics-dependent model performs the best when compared to results from LES with higher resolution. \citet{ayala_moving_2024} represent unresolved small-scale waves (ripples) as proportional to a prescribed r.m.s.\ ripple height, $\eta_{rms}'$. This follows \citet{geva_wall_2022}, who found that mean velocity profiles over young wind waves in a laboratory setting follow the fully rough log law with $\eta_{rms}'$ acting as the equivalent sand-grain roughness. These approaches all compute a globally-averaged subfilter roughness.

The subfilter wave drag model presented by \citet{aiyer_dynamic_2024}, like approaches by \citet{anderson_dynamic_2011} and \citet{yang_dynamic_2013}, employs a dynamic procedure for the subfilter wave drag. This procedure is analogous to the dynamic procedure used to define the eddy viscosity to close the subfilter stress in LES \citep{germano_dynamic_1991}. In the context of the subfilter wave drag, the subfilter wave roughness is characterized by a dynamic parameter $\alpha_w$, which is a global coefficient. Equating the total wave drag (resolved plus unresolved) at the LES (grid) filter scale and a larger test filter scale, typically twice the grid filter size, determines the value of the dynamic coefficient. Performing a dynamic procedure in this way allows for a specific subfilter drag driven by the real-time conditions rather than being a constant or predetermined value. To do so, it assumes that the dynamic parameter $\alpha_w$ is scale invariant, that is, it does not change dramatically when varying filter size; this assumption has not been previously assessed. Using a global dynamic parameter additionally assumes that the equivalent roughness does not vary significantly in space. This assumption may be tenable in spatially-homogeneous flows but deteriorates in configurations where the drag profile may change over the course of the domain (e.g., non-periodic domains or in an offshore wind farm where the wind speed characteristics change suddenly in turbine wakes, or for waves moving into a spatially varying current).

This paper develops a new subfilter drag model with a local (rather than global) dynamic parameter, proposes an alternate formulation to that described by \citet{aiyer_dynamic_2024}, and investigates the impact of changing wave conditions on MABLs. The new subfilter wave drag formulation includes the relative slip velocity between the wind and the subfilter waves to make the wave drag dynamic procedure scale invariant. The paper then compares the wave drag model to results from aggregated field data \citep{edson_exchange_2013}.
The updated local subfilter slip model is demonstrated to perform similarly to the previous model presented by \citet{aiyer_dynamic_2024} for laboratory-scale conditions and to be less grid-dependent at full-scale conditions, demonstrating improvement to the model. A parameter sweep of MABLs run with the wave drag model where wind and wave parameters are varied independently is presented to demonstrate that the relationship is more complex than a universal parameterization allows.

Section \ref{sec:waves} describes how the ocean waves are implemented in the LES and the updated subfilter wave drag model. Section \ref{sec:a_priori} presents an a priori analysis to demonstrate the model's scale invariance. Section \ref{sec:LES} outlines the computational framework used to run the simulated MABLs 
and validation of model performance in LES. Section \ref{sec:results} presents the results of the full-scale MABL study and implications for MABL surface roughness from waves. Section \ref{sec:conclusion} outlines the conclusions of this paper.

\section{Wave drag model}
\label{sec:waves}

This section describes how ocean waves are incorporated in the LES framework. Section \ref{subsec:wave_spectra} defines the ocean wave spectrum parameterization; Section \ref{subsec:wsdm} describes how the resolved wave drag is implemented in LES; and Section \ref{subsec:slip} presents the updated subfilter wave model.

\subsection{Ocean wave spectrum parameterization}
\label{subsec:wave_spectra}
Ocean waves are assumed to be one-dimensional and moving in the streamwise direction with the wind. The waves are represented with the Equilibrium wave spectrum, which determines the energy $\phi(k)$ in each wave mode $k$ 
\citep{williams_observations_2026, toba_local_1973, phillips_spectral_1985, wu_breaking_2023} :
\begin{equation}
    \label{eq:EkE}
    \phi(k) = Pu_s g^{-1/2}k^{-5/2}\exp\left[\frac{5}{4}\left(\frac{k_p}{k}\right)^2\right] \text{ ,}
\end{equation}
where $k_p$ is the peak wavenumber 
and $u_s$ is some characteristic velocity. The prefactor $P$ determines the total energy and relates to the significant wave height $H_s$:
\begin{equation}
    \label{eq:hs}
  H_s = 4\sqrt{\int\phi(P,k)\text{d}k} \text{ .}
\end{equation}
In this way, the wave spectrum is fully described by the two wave parameters $k_p$ and $H_s$. 
In the original framework of \citet{phillips_spectral_1985} considering wind and waves in equilibrium, $u_s = u_\ast$. However, the wind and waves are frequently in disequilibrium in the open ocean \citep{sullivan_large-eddy_2008, edson_exchange_2013}. The intent of this study is to examine the drag relationship that develops given prescribed wave conditions, that is, the wind adjustment to a specific wave field for a given far field wind velocity. With this in mind, $u_s$ is lumped into the coefficient $P$, and the wave field intensity is characterized through the significant wave height $H_s$ (related to the total energy) according to equation~\eqref{eq:hs} in which only the product $Pu_s$ matters. A note on the use of the term ``equilibrium'': the Equilibrium spectrum is named for the range, proposed by \citet{phillips_spectral_1985}, in which the three terms of the wave energy balance---energy input from wind, dissipation from breaking, and transfer between scales from nonlinear wave-wave interactions---are in equilibrium with each other, represented by the $k^{-5/2}$ relationship in the tail \citep{williams_observations_2026,deike_mass_2022,wu_breaking_2023}. Here the Equilibrium wave spectrum shape is used, and the wave conditions $H_s$ and $k_p$ are varied to define a wave field independent of the far field wind condition above it (in other words, direct description rather than indirect description) and diagnose the resulting drag measurable through the adjustment of the wind field near the wave surface. In this approach, we will experiment with waves conditions that might be far from the ones expected for a given wind amplitude.

The instantaneous surface elevation for each wave mode $\eta_n(k)$ is calculated as 
\begin{equation}
  \label{eq:eta}
  \eta_n(k_n,x,t) = a_n\cos\left(k_nx - \omega_n t + \theta_n\right) \text{ ,}
\end{equation}
where the amplitude $a_n = \sqrt{2\phi(k_n)\text{d}k_n}$ is the wave mode amplitude, the frequency is determined by the dispersion relation for gravity waves $\omega_n = \sqrt{gk_n}$, and the phase shift $\theta_n$ is assigned randomly for each wave mode. 
$\text{d}k_n$ is determined by the resolution of the vector $k$. The wave drag model, detailed below, does not require explicit computation of the filtered wave height $\widetilde{\eta}$ but instead of $\frac{\partial\widetilde{\eta}}{\partial x}$, which can be determined analytically from equation~\eqref{eq:eta}. 

\begin{figure}[htbp]
  \centering
    \includegraphics[width=0.8\textwidth]{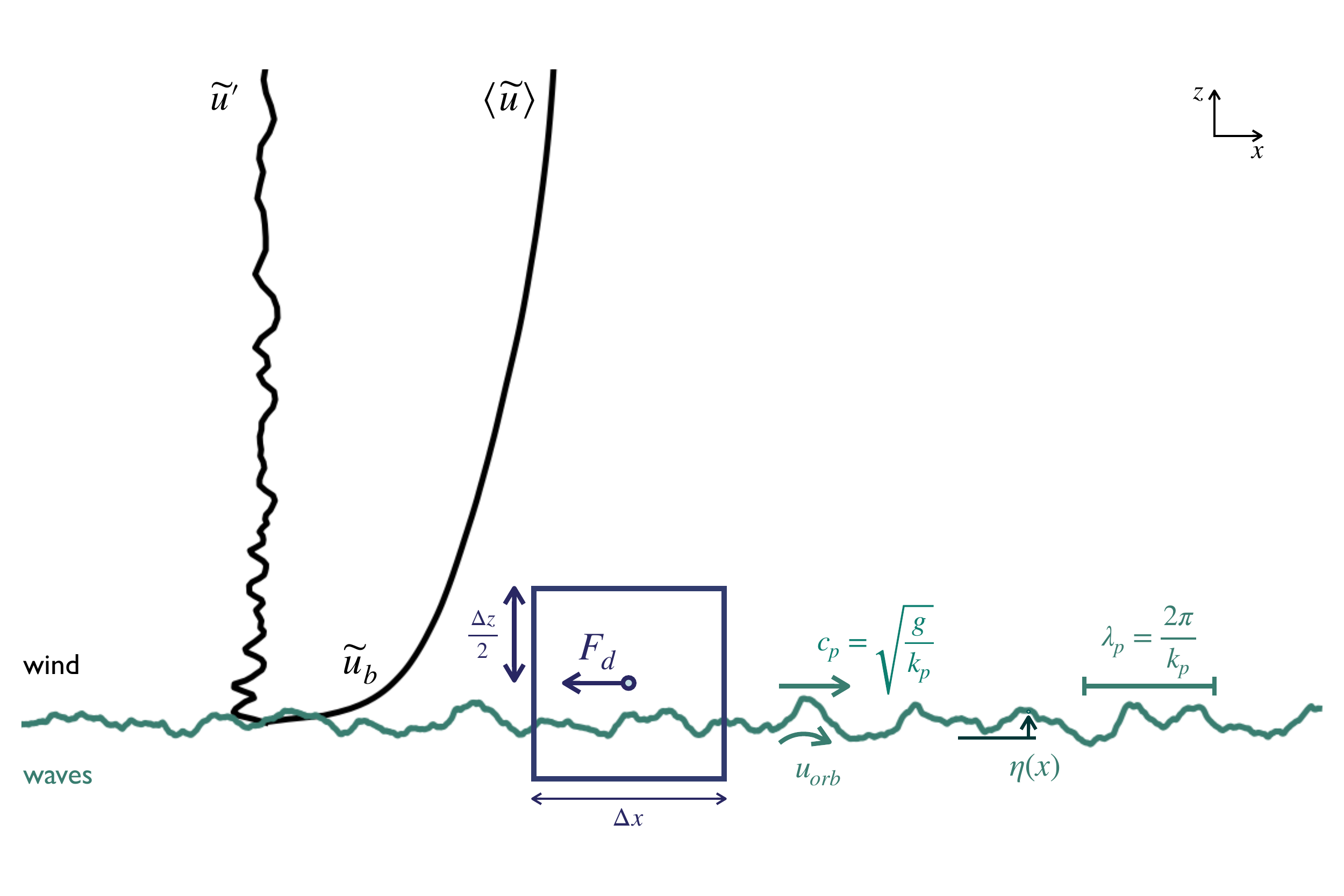}
  \caption{Sketch of the configuration showing a turbulent Marine Atmospheric Boundary Layer over a one-dimensional spectrum of ocean waves. An example of the bottom cell in the Large Eddy Simulation with the applied drag force from the waves is shown as well. Wave parameters are in green: peak wavelength $\lambda_p$, peak phase speed $c_p$, instantaneous surface height $\eta$, and orbital velocity $u_{orb}$. Wind parameters are in black: an example of the mean filtered velocity profile $\langle\widetilde{u}\rangle$, the filtered fluctuation $\widetilde{u'}$, and the filtered velocity at the bottom $\widetilde{u}_b$.}
  \label{fig:sketch}
\end{figure}

\subsection{Wave drag formulation}
\label{subsec:wsdm}
The wave drag model to represent the drag from a spectrum of waves was initially proposed in \citet{aiyer_dynamic_2024}. The drag contribution from waves larger than the streamwise grid spacing $\Delta_x$ is calculated for each wave mode by considering the first grid point above the sea surface as a control volume and applying a hydrodynamic drag force proportional to the incoming momentum flux onto the front surface area of the wave \citep{aiyer_sea_2023}. An example cell is shown in Figure \ref{fig:sketch}.

Each wave mode's amplitude is defined by the wave spectrum, discussed in Section \ref{subsec:wave_spectra}. The wave mode has filtered wave height $\widetilde{\eta}_n = \widetilde{\eta}_n(x, t, k_n)$ defined from equation~\eqref{eq:eta}; the filtered local sea surface elevation is described by the sum of linear plane waves, each with a random phase $\theta_n$:
\begin{equation}
  \widetilde{\eta}(x,t) = \sum_{k_n=k_{min}}^{2\pi/\Delta_x} a_n\cos (k_nx - \omega_n t + \theta_n) \text{ .}
\end{equation}
$k_{min}$ is determined based on the streamwise domain size and is small enough such that the drag contribution from larger waves would be negligible. 

The effect of the waves is expressed as a force per unit mass on the right hand side of the momentum equation for a single one-dimensional wave mode (denoted by subscript $[\cdot]_n$) moving in the streamwise direction, separated into regimes based on whether the wave is a wind wave (slower than the wind) or a swell wave (faster than the wind):
\begin{equation}
  \label{eq:res_waves}
  F_{d,n} = F_{d,n}(x, y, t, k_n) = \begin{cases}
    C_{d,n} \frac{1}{\Delta_z} \widetilde{u}_b (\widetilde{u}_b - c_n) \frac{\partial\widetilde{\eta}_n}{\partial x} \mathcal{H}\left\{\frac{\partial\widetilde{\eta}_n}{\partial x}\right\} & \text{if } \widetilde{u}_b > c_n \\
    \beta(k_n)\frac{(a_nk_nu_\ast)^2}{2} & \text{if } \widetilde{u}_b < c_n \text{, } \frac{c_n}{u_\ast} > 25 \text{, and } k_n < 2k_p \\
    0 & \text{otherwise.}
    \end{cases}
\end{equation}
In equation~\eqref{eq:res_waves}, $\Delta_z$ is the local vertical grid size, $\widetilde{u}_b = \widetilde{u}_b(x, y, t)$ is the filtered streamwise velocity at the first (bottom) grid point above the sea surface, and $c_n = \sqrt{g/k_n}$ is the wave mode phase velocity defined by the dispersion relation for gravity waves. $\mathcal{H}\{x\}$ is the Heaviside function, which ensures that the force is only applied when the flow is incident on the wave frontal area. The scale- and amplitude-dependent local drag coefficient is written as $C_{d,n} = a_nk_n/(1+6(a_nk_n)^2)$ and is formulated such that the energy input rate for a single mode $S_{in, \text{ LES}} = F_{d,n}\Delta_z c_n$ scales similarly to the sheltering parameterization proposed by \citet{jeffreys_formation_1925}, which scales quadratically with the steepness $S_{in,\text{ Jeffreys}} \sim (ak)^2$, modulated by a prefactor $1/(1+Q(ak)^2)$ according to \citet{belcher_wave_1999}, where $Q$ is a coefficient fit to their data \citep{aiyer_sea_2023}.  $\beta(k_n) = 25 - \frac{c_n}{u_\ast}$ is the wave energy input rate for swell waves, parameterized through a fit to phase-resolved simulations \citep{cao_numerical_2021}; this empirical model was shown to be valid for longer waves with $c_n/u_\ast > 25$. Since the drag model is local, there is a possibility that a locally low wind speed would result in smaller-wavelength waves activating the swell model, so a further restriction that $k_n < 2k_p$ is added. The drag for swell waves does not depend on the local phase of the wave mode so is not phase-aware.

The total drag force imparted on each grid cell is calculated as a sum of the individual wave mode contributions:
\begin{equation}
  F_d(x,y,t) = \sum_{k_n = k_{min}}^{2\pi/\Delta_x} F_{d,n}(x,y,t,k_n) \text{ .}
\end{equation}
Note that the total drag force is applied in the negative $x$ direction (opposing the streamwise flow) as denoted in the sketch (Figure \ref{fig:sketch}), so wind waves contribute positively to the total drag and swell waves contribute negatively. The resulting wave drag model is phase-aware, vertically unresolved, and horizontally resolved for wave modes that are larger than the LES filter size. 
Misaligned wind and waves are not accounted for by the model so are left to future work.

\subsection{Dynamic subfilter wave drag model with slip}
\label{subsec:slip}
Subfilter wave effects---drag from waves smaller than the LES grid size---are captured with an equilibrium log-law model for neutrally-stratified flow \citep{anderson_dynamic_2011, yang_dynamic_2013}:
\begin{equation}
    \label{eq:sf_stress}
  \widetilde{\tau}_w = \left[\frac{\kappa \left(\widetilde{u}_b-\widetilde{u}_{orb}\right)}{\log\left(\frac{\Delta_z/2}{z_{0,\Delta}}\right)}\right]^2 \text{ ,}
\end{equation}
where $\widetilde{u}_{orb} = \sum_na_n\omega_n\cos(k_nx - \omega_nt + \theta_n)$ is the surface wave orbital velocity, $\kappa$ is the von Karman constant, and $\Delta_z/2$ is the center of the bottom-most grid point. The equivalent subfilter wave roughness $z_{0,\Delta}$ is defined based on the r.m.s.\ of the wave height fluctuation $\sigma_\eta^\Delta = \left(\widetilde{\eta^2} - \widetilde{\eta}^2\right)^{1/2}$ and a dynamic parameter $\alpha_w$:
\begin{equation}
  z_{0,\Delta} = \left[z_{0,s}^2 + \left(\alpha_w B \sigma_\eta^\Delta\right)^2\right]^{1/2} \text{ ,}
\end{equation}
where the smooth wall roughness $z_{0,s} = 0.11\nu/u_\ast$ \citep{fairall_bulk_1996} is very small in magnitude compared to most subfilter wave roughnesses but prevents the total equivalent subfilter roughness from going to zero. We will discuss the factor $B$ below.
A spatially-averaged total $u_\ast$ combining resolved and subfilter effects is computed on-the-fly each time step (see equation~\eqref{eq:us_les} and discussion in Section \ref{subsec:ustar}) and used for calculation of $z_{0,s}$ as well as to set the condition for swell modes in the resolved drag model (equation~\eqref{eq:res_waves}). The magnitude of $z_{0,s}$ ($\sim\mathcal{O}(10^{-6})$ m) is much smaller than that of typical $z_{0,\Delta}$ ($\sim\mathcal{O}(10^{-3}\text{ to }10^{-2})$ m). A dynamic procedure is used to calculate the coefficient $\alpha_w$: the wave drag is calculated for wave parameters at the filter scale $\Delta$ and a test filter $\widehat{\Delta} = 2\Delta$. The dynamic parameter $\alpha_w$ is determined by requiring that the total drag be the same at both scales, that is, 
\begin{equation}
  \label{eq:dynamic_drag}
  \widetilde{F}_d \Delta_z + \widetilde{\tau}_w = \widehat{\widetilde{F}}_d\Delta_z + \widehat{\widetilde{\tau}}_w \text{ ,}
\end{equation}
where $\:\widehat{\widetilde{\cdot}}\:$ denotes double-filtering at the test filter; 
the waves are filtered using a Gaussian filter. The construction of the subfilter wall stress in equation~\eqref{eq:sf_stress} and the dynamic procedure in equation~\eqref{eq:dynamic_drag} differs from that in previous approaches \citep{anderson_dynamic_2011, yang_dynamic_2013, aiyer_dynamic_2024} in two ways. The first is that the bottom velocity used in both the grid- and test-filtered drag quantities (resolved and unresolved) is always $\widetilde{u}_b$, the grid-filtered surface velocity. This is to ensure that the wave drag responding to a specific wind speed is determined. The second is that the logarithmic term in the denominator of the subfilter wall stress only considers the distance above the mean surface. Previous approaches computed a global dynamic coefficient for the whole domain at each time step, that is, considering each side of equation~\eqref{eq:dynamic_drag} averaged over the entire surface. Both of these changes are made because they make the local solution of $\alpha_w$ more computationally robust. Without these changes, the dynamic procedure to solve for $\alpha_w$ frequently has no valid solution; with the changes, instances where there is no root for $\alpha_w$ are limited to when $\widehat{\widetilde{F_d}}$ is negligibly small, that is, no resolved drag at the test filter, which is expected behavior and indicates severe under-resolution of wavelength of drag-imparting waves. 

The subfilter wave drag model originally proposed by \citet{aiyer_dynamic_2024} does not include the factor $B$ so is equivalent to $B = 1$. However, the model for the subfilter stress was developed for stationary rough surfaces \citep{anderson_dynamic_2011}, so a new formulation is proposed to add a nondimensional slip factor:
\begin{equation}
  \label{eq:slipB}
  B = \frac{\widetilde{u}_b - c_{sf}}{\widetilde{u}_b + c_{sf}} \text{ .}
\end{equation}
In equation~\eqref{eq:slipB}, a characteristic subfilter wave phase speed $c_{sf}$ is calculated from a characteristic subfilter wavenumber weighted by the wave energy spectrum:
\begin{equation}
  \label{eq:ksf}
  k_{sf} = \frac{\int_{k_{\Delta_x}}^{k_{\lambda_c}} k^m\phi(k)dk}{\int_{k_{\Delta_x}}^{k_{\lambda_c}} k^{m-1}\phi(k)dk} \text{ ,}
\end{equation}
where $k_{\lambda_c}$ is the wavenumber associated with capillary length scale (smallest wave), $k_{\Delta_x} = 2\pi/\Delta_x$ is the cutoff wavenumber defined by the LES (grid) filter size, and $m\geq 1$ (the value of $m$ is discussed in Section \ref{sec:a_priori}). The characteristic subfilter wave speed $c_{sf} = \sqrt{g/k_{sf}}$ is determined by the dispersion relation for gravity waves. Additionally, the dynamic coefficient is made local such that $\alpha_w$ = $\alpha_w(x, y, t)$ to accommodate spatial heterogeneity, and equation~\eqref{eq:dynamic_drag} is balanced everywhere along the surface. 
The form of $B$ ensures that the subfilter wave roughness scales correctly with the slip velocity. As $c_{sf}$ goes to 0, $B \rightarrow$ 1, which recovers the limit of a stationary surface. As the velocities become equal, on the other hand, $B\rightarrow 0$, and the equivalent subfilter wave roughness disappears. Furthermore, $B$ is restricted to $0 \leq B \leq 1$. The physical impact of the formulation for $k_{sf}$ in equation~\eqref{eq:ksf} is discussed in Appendix~\ref{app:ksf}.

The dynamic parameter $\alpha_w$ is found using a bisection root-finding algorithm following \citet{aiyer_dynamic_2024} at every cell on the bottom surface of the LES domain. There is a discontinuity in $\tau_w(\alpha_w)$ that results from the denominator going to zero if $z_{0,\Delta} = \Delta_z/2$; the correct value of $\alpha_w$ should always be smaller than the location of the discontinuity since the LES grid should be set up such that $\Delta_z/2 > H_s$. With the modifications to the subfilter stress model described above, the instances in which there is no root are limited to those in which there is not enough drag resolved. $\alpha_w$ in this case is set to the global average of valid values. 

Adding a factor to account for the slip velocity acknowledges the dispersion relation for gravity waves in the construction of the subfilter wave drag model: because waves of different sizes travel at different speeds, their contribution to the roughness should take into account their velocity relative to the wind. Defining the subfilter roughness without slip is akin to defining the wave roughness solely based on $H_s$---starting with Charnock, roughness parameterizations have acknowledged the influence of the wind velocity in wave roughness \citep{Charnock1955}. 
The intent behind the slip model is similar to that of the wave-kinematics-dependent approach described in \citet{yang_dynamic_2013}---that is, to incorporate the propagation speed of the subfilter waves. However, the fundamental construction differs: where \citet{yang_dynamic_2013} superpose contributions to the roughness from discrete wave modes, this model seeks to identify a representational subfilter wave mode. Critically, this model also computes the subfilter wave drag locally rather than through a global average. Accounting for the slip allows for the model to be implemented fully locally because the roughness characteristics may differ for a heterogeneous wind.

\subsection{Resolution constraints with the wave drag model}
\label{subsec:res}
There are a number of competing factors constraining the LES grid resolution, listed here for ease of reproducibility. First, it is important that the peak wavelength is horizontally resolved by the test filter in order to have resolved drag and for the dynamic procedure to work, so $\Delta_x > \frac{\lambda_p}{4} = \frac{\pi}{2k_p}$. Second, the waves should be vertically confined to the bottom half of the first grid cell, so $\Delta_z > 2H_s$. Lastly, the grid ratio for wall-modeled LES should be such that $\Delta_x/\Delta_z > 1$ \citep{Piomelli2008, kawai_wall-modeling_2012}.

\section{Dynamic parameter scale invariance: an a priori study}
\label{sec:a_priori}

\begin{figure}[htbp]
  \centering
    \includegraphics[width=0.9\textwidth]{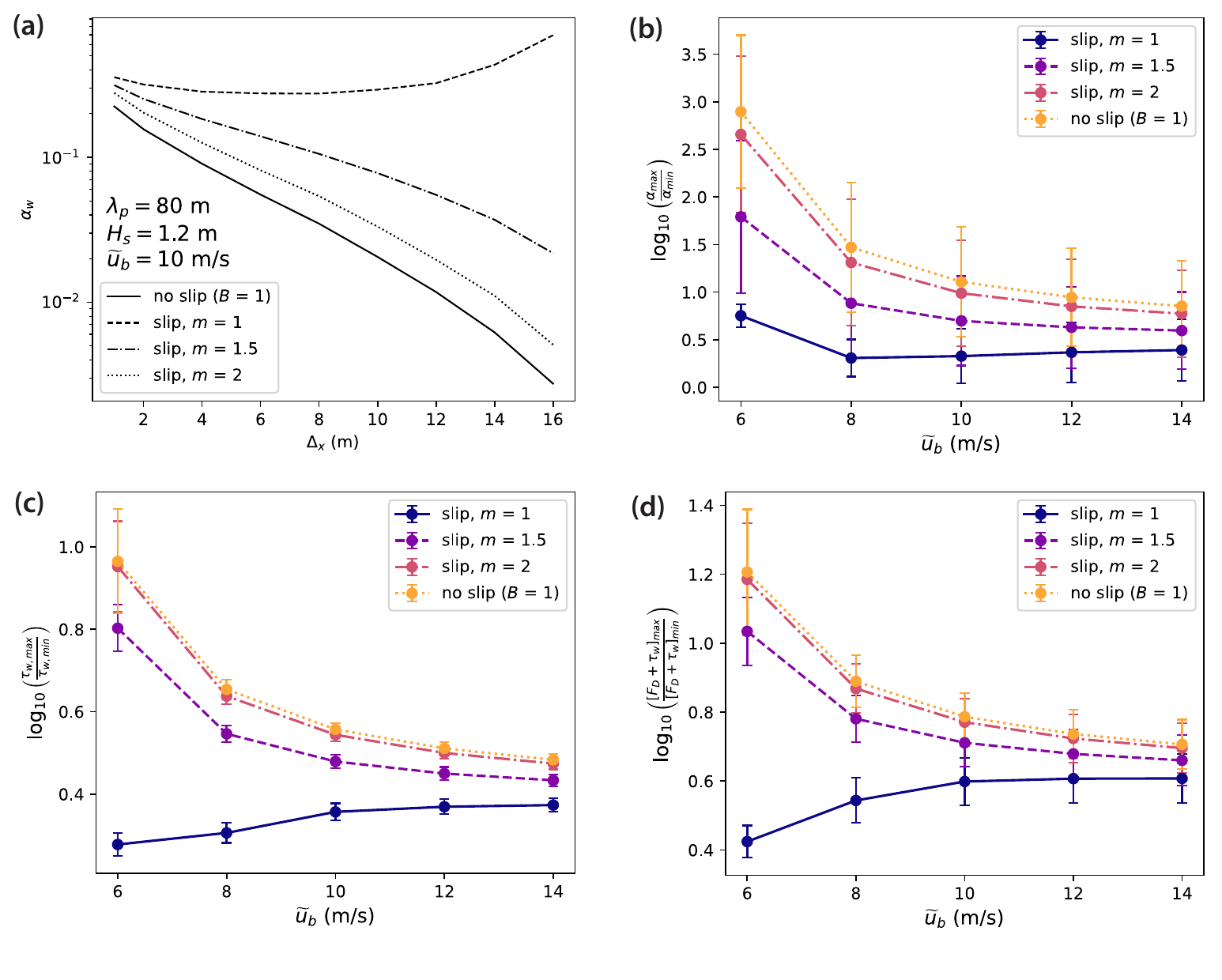}
  \caption{(a) Values of $\alpha_w$ calculated in a priori analysis for three subfilter wave drag models over grid size $\Delta_x$ for one set of wind and wave conditions. (b) The difference in order of magnitude between the maximum and minimum value of $\alpha_w$ for $\Delta_x$ from 1 to 16 m is calculated for each wave condition; the mean (circles) and standard deviation (error bars) are plotted over the surface velocity. The same is done for the computed subfilter wave drag (c) and total wave drag (d).}
  \label{fig:scalevar}
\end{figure}

The use of a dynamic procedure to calculate the subfilter roughness coefficient $\alpha_w$ assumes that $\alpha_w$ is scale invariant at relevant filter sizes. An a priori analysis was conducted to compare the scale variance of different subfilter wave models 
by isolating the iterative calculation of $\alpha_w$ in a standalone script given constant wind and wave conditions and changing only the filter size and subfilter wave model. $\widetilde{\eta}$ and $\widetilde{u}_s$ were taken as the r.m.s.\ of a randomly-generated virtual wave field based on the wave spectrum. Using these predetermined wind and wave conditions, $\alpha_w$ was calculated to satisfy equation~\eqref{eq:dynamic_drag} between the LES filter $\Delta_x$ and a test filter $2\Delta_x$. This analysis was conducted for 20 different wave conditions spanning $\lambda_p = [40, \:80, \:120, \:160, \:200]$ m and $H_s = [0.7, \:1.2, \:1.7, \:2.2]$ m 
as well as five surface wind velocities $\widetilde{u}_b = [6, \:8, \:10, \:12, \:14]$ m/s for a total of 100 unique conditions. An example of the result for one wave condition is displayed in Figure \ref{fig:scalevar}a comparing calculated values of $\alpha_w$ for the three subfilter wave drag models (i.e., formulations for $B$ in equation~\eqref{eq:slipB}). The no slip model causes $\alpha_w$ to shrink by several orders of magnitude between $\Delta_x = 1$ m and $\Delta_x = 16$ m; the slip model with $m = 1$ increases slightly in magnitude but stays relatively constant through $\Delta_x = 12$ m. The slip model with $m = 2$ behaves similarly to the no slip model. This is because the characteristic subfilter wavenumber $k_{sf}$ is large enough that the slip factor $B$ approaches unity---the no slip model is equivalent to the slip model as $m \rightarrow \infty$.

Figure \ref{fig:scalevar}b aggregates the results from the 20 wave conditions and plots the mean (circles) and standard deviation (error bars) of $\log\left(\frac{\alpha_{w,max}}{\alpha_{w,min}}\right)$ between $\Delta_x = $ 1 and 16 m as a function of bottom wind velocity, separated by model (line color/style). Plotting this quantity shows the orders of magnitude spanned by values of $\alpha_w$ as the horizontal grid spacing changes, that is, to quantify the scale variance over $\Delta_x$. The slip model with $m = 1$ has consistently low variation in $\alpha_w$ as the grid size changes for tested wave conditions, particularly at lower wind speeds---when the bottom velocity is 6 m/s, the magnitude of $\alpha_w$ varied by 3.5 orders of magnitude for the no slip case but less than one order of magnitude for the no slip model with $m = 1$. The same analysis is conducted to plot the orders of magnitude spanned by subfilter wave drag (Figure \ref{fig:scalevar}c) and total wave drag (Figure \ref{fig:scalevar}d) to demonstrate how the impact of variation in the scale of $\alpha_w$ propagates. Although the variation in scale in the total wave drag is smaller than that of $\alpha_w$ for the no slip model and slip models with higher $m$, variation of over one order of magnitude is still significant for low wind speed. However, for all quantities, the variation is lower with $m = 1$ compared to the other models, especially for lower wind speeds. Most of the variation comes from $\Delta_x$ at the extreme ends of tested resolutions. 

The analysis only considers variations in $\alpha_w$ up to $\Delta_x = 16$ m; the dynamic procedure requires the presence of resolved drag at both the grid and test filter scales, so coarser grids would produce non-physical results. This a priori study shows that adding the slip factor in equation~\eqref{eq:slipB} makes the dynamic parameter vary drastically less with filter size in the relevant ranges, making the assumption of scale invariance that is central to performing the dynamic procedure for subfilter wave drag valid.

\section{Simulations of Marine Atmospheric Boundary Layer}
\label{sec:LES}
This section discusses the use of the scale invariant, spatially-varying subfilter wave drag model in simulations of full-scale MABLs.

\subsection{LES framework}
\label{subsec:methods}
LES calculations are performed using NGA, a structured, finite difference, low Mach number flow solver using second-order centered schemes for spatial discretization and a second-order, iteratively-implicit temporal integration scheme \citep{Desjardins2008, MacArt2016}. The filtered Navier-Stokes equations in the incompressible limit describe the air velocity field:
\begin{equation}
    \nabla \cdot \widetilde{\mathbf{u}} = 0 \text{ ,}\hspace{2em}\text{and} \hspace{3em}
    \frac{\partial\widetilde{\mathbf{u}}}{\partial t} + \widetilde{\mathbf{u}} \cdot \nabla\widetilde{\mathbf{u}} = - \frac{1}{\rho} \nabla \widetilde{p} + \nabla\cdot\widetilde{\underline{\underline{\boldsymbol{\sigma}}}} + \widetilde{\mathbf{F}}_d \text{ .}
\end{equation}
These equations are discretized on a Cartesian grid $(x,y,z)$, where $x$ and $y$ are the streamwise and spanwise directions, respectively, and $z$ is the vertical coordinate. Here, $\widetilde{\mathbf{u}} = (\widetilde{u},\widetilde{v},\widetilde{w})$ is the velocity vector in the corresponding directions, where the tilde denotes variables filtered on the LES grid. $\widetilde{\mathbf{F}}_d$ is the resolved drag force representing the effects of the waves, discussed in Section \ref{subsec:wsdm}, applied in the streamwise direction to the bottom-most layer of cells. $\widetilde{\sigma}_{ij} = 2\nu \widetilde{S}_{ij} + \widetilde{\tau}_{ij}^d$ is the total deviatoric stress, where $\nu$ is the molecular viscosity, $\widetilde{S}_{ij}$ is the resolved strain rate tensor, and $\widetilde{\tau}_{ij}^d$ is the subfilter stress (SFS) tensor. The SFS tensor is modeled using a Lilly-Smagorinsky type subfilter viscosity model $\widetilde{\tau}_{ij}^d = 2\nu_T \widetilde{S}_{ij}$, where the subfilter viscosity is computed using the Anisotropic Minimum Dissipation (AMD) model \citep{abkar_large-eddy_2017, rozema_minimum-dissipation_2015}. 

The simulations are run with a constant mass flux rather than a prescribed pressure gradient, allowing for the MABL to develop without an imposed global friction velocity. Instead, a bulk velocity is predetermined. The flow is driven through a pressure gradient $\nabla\widetilde{p}$ that is calculated at each time step to keep the bulk mass flux of the LES constant \citep{lenormand_large_2000, quadrio_does_2016}. Prescribing a pressure gradient sets $u_\ast$; since the behavior of interest is the drag relationship at the sea surface, a parameter sweep by independently changing wind and wave parameters is most effectively conducted in a computational framework that does not prescribe the global friction velocity. A discussion of and comparison to pressure gradient-driven MABLs can be found in Appendix~\ref{app:pgrad}. 

The bottom boundary of the LES represents the ocean waves and is as described in Section \ref{sec:waves}. The top of the domain has a free-slip boundary. The boundary layer is allowed to develop until statistically stationary, around time $t \approx 12H/u_\ast$, where $H = L_z = 1000$m is the eventual boundary layer height (exact boundary-layer time scale varies based on measurement of $u_\ast$---see Section \ref{subsec:ustar} for more details). Mean velocity, second-order statistics, and friction velocity were all demonstrated to be statistically stationary at this time. Boundary layers were run for another 10,000 s to generate statistics for analysis, discussed in Section \ref{sec:results}.
A comparison of cases run with constant mass flux and those driven by a constant pressure gradient can be found in Appendix~\ref{app:pgrad}. A vertical grid resolution study ($\Delta_z$ = 3, 4, and 5 m) demonstrated minimal impact of changes in vertical grid size: a 4.4\% change in mean streamwise velocity at 10 m above the sea surface and 1.6\% change in $U_{20}$. There is some sensitivity in the second order statistics to vertical grid resolution, but resulting $u_\ast$ calculated from the vertical momentum flux varies by less than 3\% between $\Delta_z$ = 3 m and $\Delta_z$ = 4 m. 

\begin{figure}[htbp]
    \centering
    \includegraphics[width=0.5\linewidth]{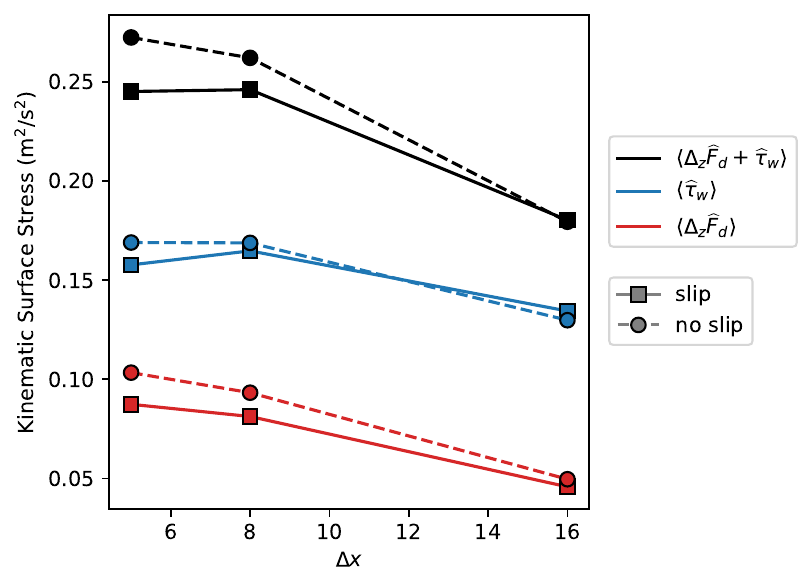}
    \caption{
    Resolved, subfilter, and total stress at wind-wave boundary from no slip and slip subfilter wave drag models (both local) as a function of grid size for the horizontal grid convergence study run in LES. 
    }
    \label{fig:hgrid}
\end{figure}

\afterpage{\afterpage{
\begin{table}[t!]
  \caption{Parameters of analyzed MABL cases, separated and listed by varied parameter ($U_{bulk}$, $H_s$, and $k_p$, respectively). The same middle case, denoted with asterisks, is used for all three swept parameters. Friction velocity $u_\ast$ for the wave age $c_p/u_\ast$ is computed as an average of the vertical momentum flux $\langle u'w'\rangle$ from 0-100 m 
  (see section \ref{subsec:ustar} for a discussion on measuring $u_\ast$). Frequently-used nondimensional wave parameters (the significant wave height $k_pH_s$ and wave age $c_p/u_\ast$ or $c_p/U_{10}$) are also included, along with the wind velocity at the top of the domain $U_\infty$ and the mean value of $\alpha_w$. 
  All MABLs in this study are neutral. }
  \centering
  \footnotesize
  \setlength{\tabcolsep}{3.5pt}
    \begin{tabular}{ c | *{6}{c} | *{4}{c}}
      \hline 
      & \multicolumn{6}{c|}{Inputs (set parameters)} & \multicolumn{4}{c}{Results from simulation} \\
      ID & $U_{bulk}$ (m/s) & $H_s$ (m) & $k_p$ (m$^{-1}$) & $\lambda_p$ (m) & $c_p$ (m/s) & $k_pH_s$ & $c_p/u_\ast$ & $c_p/U_{10}$ & $U_\infty$ (m/s) & mean $\alpha_w$ \\
      \hline
      U15 & 15 & 1.2 & 0.0785 & 80 & 11.18 & 0.094 & 21.92 & 1.20 & 15.9 & 0.21 \\
      Mid & 18* & 1.2* & 0.0785* & 80* & 11.18* & 0.094* & 18.79* & 1.01* & 19.5 & 0.23 \\
      U21 & 21 & 1.2 & 0.0785 & 80 & 11.18 & 0.094 & 15.44 & 0.88 & 22.6 & 0.21 \\
      U24 & 24 & 1.2 & 0.0785 & 80 & 11.18 & 0.094 & 13.18 & 0.76 & 25.9 & 0.21 \\
      \hline
      Hs04 & 18 & 0.4 & 0.0785 & 80 & 11.18 & 0.157 & 25.18 & 0.85 & 18.8 & 0.0028 \\
      Hs20 & 18 & 2.0 & 0.0785 & 80 & 11.18 & 0.031 & 16.44 & 1.19 & 19.7 & 0.76 \\
      \hline
      Kp04 & 18 & 1.2 & 0.0393 & 160 & 15.80 & 0.047 & 28.21 & 1.34 & 19.3 & 0.047 \\
      Kp16 & 18 & 1.2 & 0.157 & 40 & 7.91 & 0.188 & 13.31 & 0.86 & 19.6 & 0.71 \\
      \hline
    \end{tabular}
  \label{tab:param_sweep}
\end{table}
\afterpage{
\begin{figure}[t!]
    \centering
    \includegraphics[width=0.9\linewidth]{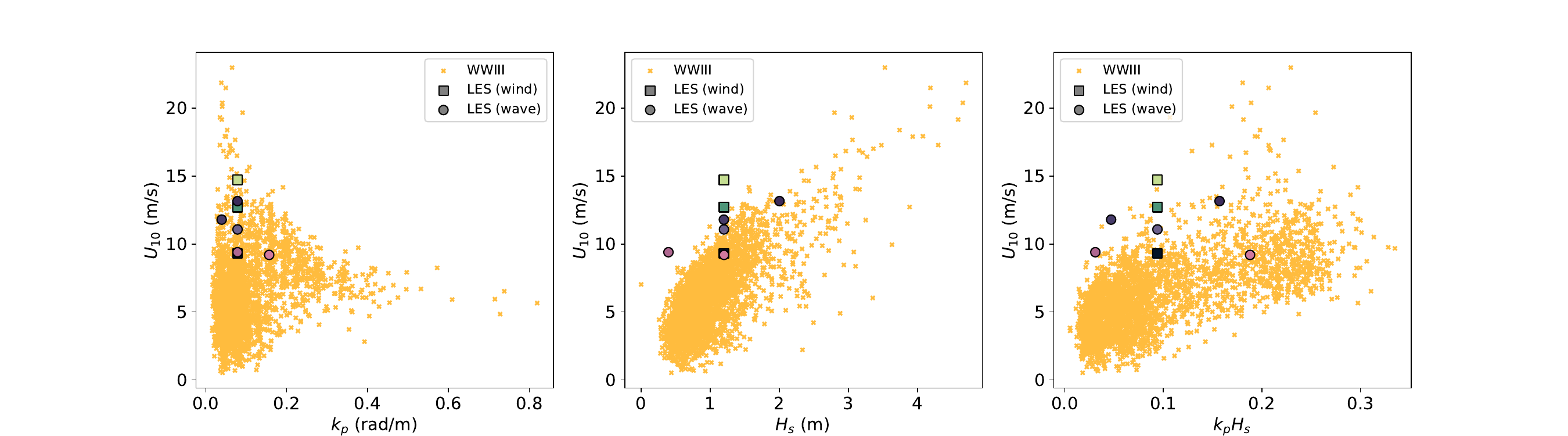}
    \caption{
    Climatology data generated through WAVEWATCH III and the JRA55-do reanalysis dataset \citep{zhou_sea_2023, the_wavewatch_iii_development_group_ww3dg_user_2019, tsujino_jra-55_2018} off the coast of New Jersey (small yellow crosses) compared to wind and wave conditions in the LES cases run as part of the parameter study. Squares correspond to cases sweeping the bulk velocity and circles correspond to cases sweeping wave condition. Shape and color of the LES cases match those shown in Figure \ref{fig:trends}. The climatology data shown here include misaligned conditions. }
    \label{fig:climatology}
    \vspace{1.5em}
    \includegraphics[width=\textwidth]{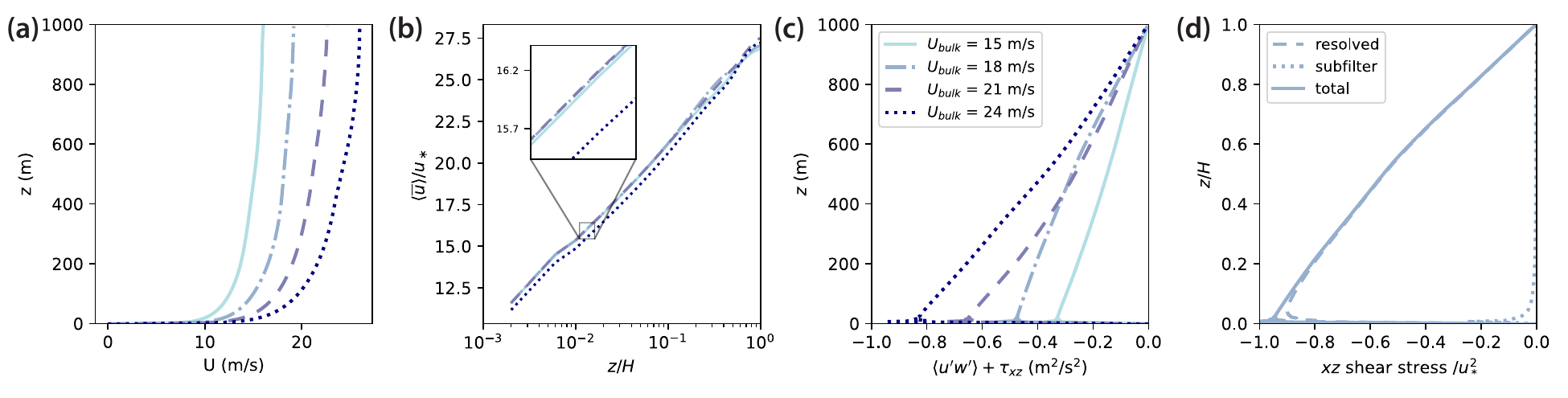}
    \caption{
    Mean velocity (left, center left), total vertical momentum flux (center right) profiles from LES of MABLs with varied bulk velocity. The right panel shows the contributions of resolved and subfilter shear stress ($xz$) for the `Mid' case. } 
    \label{fig:profiles}
\end{figure}
\afterpage{
\begin{figure}[t!]
  \centering
  \includegraphics[width=0.9\textwidth]{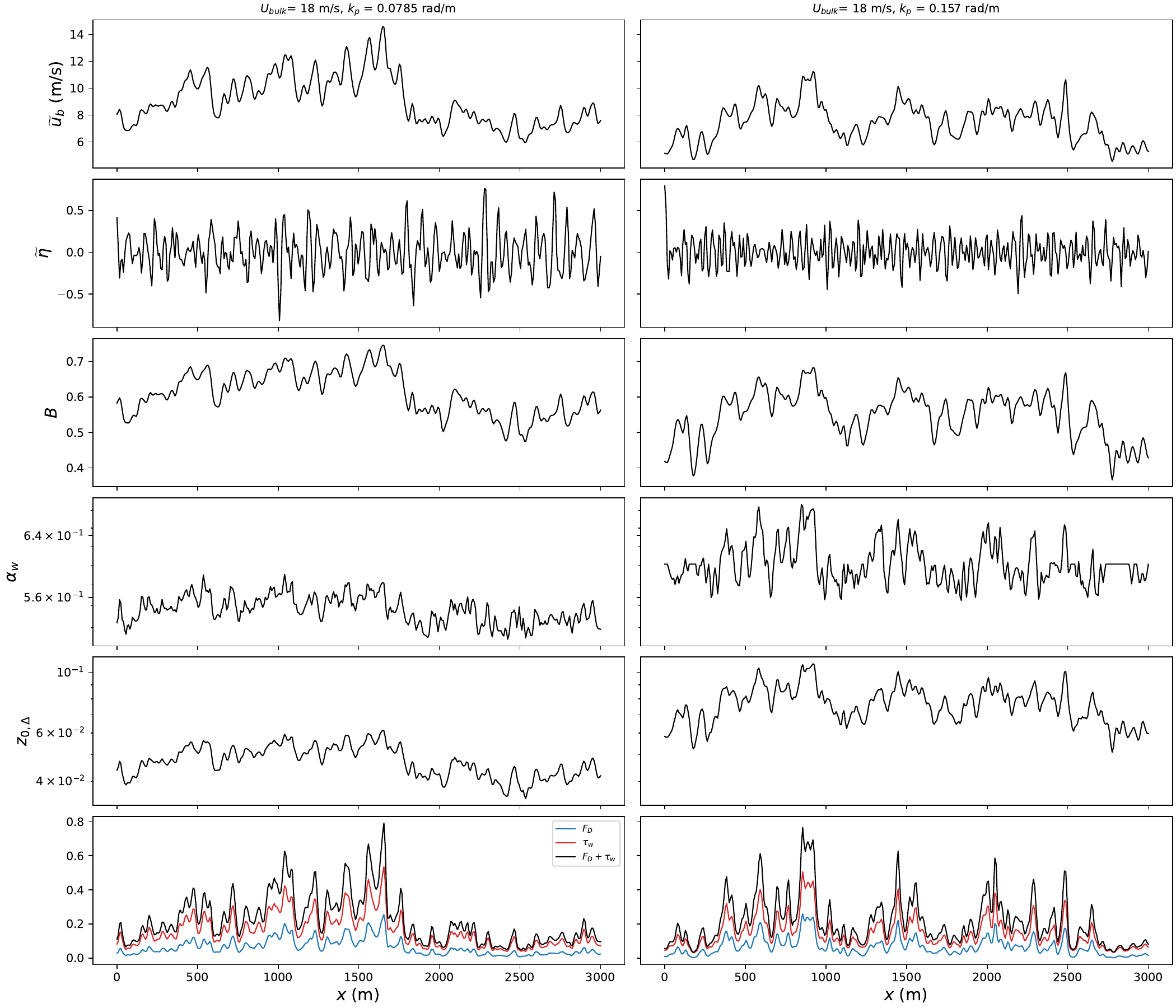}
  \caption{
  Example quantities evaluated at the bottom of the LES domain at a single time and spanwise location as a function of streamwise direction $x$ for cases Mid (left) and Kp16 (right) described in Table \ref{tab:param_sweep}. From top to bottom: filtered velocity at the bottom-most cell $\widetilde{u}_b$, filtered wave height $\widetilde{\eta}$, slip coefficient $B$, dynamic parameter $\alpha_w$, subfilter wave roughness $z_{0,\Delta}$, and kinematic surface stress, which plots resolved drag $F_D$, subfilter wave drag $\tau_{w}$, and the total wave drag $F_D + \tau_w$ together.
  } 
  \label{fig:xyplane}
\end{figure}}}}}

\subsection{Evaluating performance of subfilter slip wave model in LES}
\label{subsec:validation}
A horizontal grid study was performed in order to evaluate the behavior of the slip subfilter wave drag model in comparison to the version with no slip. Full-scale, neutral Marine Atmospheric Boundary Layers (MABLs) are generated in a streamwise- and spanwise-periodic domain with $L_x \times L_y \times L_z = 3000\:\text{m} \times 1024\:\text{m} \times 1000\:\text{m}$. The grid is uniform in the horizontal directions, for which resolutions were studied: $\Delta_x \times \Delta_y = 5 \text{ m} \times 4 \text{ m}$, $\Delta_x \times \Delta_y = 8 \text{ m} \times 8 \text{ m}$, and $\Delta_x \times \Delta_y = 16 \text{ m} \times 16 \text{ m}$. The vertical grid starts with $\Delta_{z,min} = \Delta_x/2 = 3.8$ m at the bottom of the domain and is uniform until a height of $200$ m, after which it is stretched linearly at a 3\% rate until the maximum cell size of $\Delta_{z,max} = 15$ m is reached (at around 550 m) and kept constant until the top of the domain. The flow is initialized at a constant velocity $U_{bulk} = 18$ m/s with fluctuations in $\widetilde{u}$, $\widetilde{v}$, and $\widetilde{w}$ included in the bottom $20$\% of the domain to facilitate the transition to turbulence. The wave field has $H_s$ = 1.2 m and $k_p$ = 0.0785 rad/m.

Cases were run with local drag and dynamic procedure comparing slip ($m =1$) and no slip ($B=1$) construction of the subfilter wave drag model. The mean kinematic surface stresses from this study are shown in Figure \ref{fig:hgrid}. As expected, the resolved component of the surface stress increases with the resolution as a larger portion of wave modes are resolved by the grid. The coarsest case, $\Delta_x = 16$ m, is under-resolved, as demonstrated by the small resolved drag. For the model with slip, the subfilter wave drag decreases as $\Delta_x$ goes from 8 to 5 m, which results in a similar mean total surface stress. The no slip model has slightly increasing mean subfilter stress, resulting in an increase in total stress between 5 and 8 m grids, neither of which is the correct behavior. This demonstrates that adding the slip component to the subfilter wave stress improves the wave drag model's response to changes in horizontal resolution where a decrease in subfilter effects is expected for increasing resolution. Figure \ref{fig:hgrid} further indicates that the subfilter wave drag has higher magnitude than the resolved component. This is compatible with findings by \citet{makin_drag_1995} that a large portion of form drag comes from waves with wavelength smaller than 7 m and emphasizes the importance of the subfilter wave drag model. In addition to this horizontal grid study, a validation case was run in laboratory-scale conditions to compare LES performance to the model presented by \citet{aiyer_dynamic_2024} and is described in Appendix~\ref{app:aiyer_validation}.

\subsection{Wind and wave parameter sweep setup}


The goal of the parameter sweep is to investigate the impacts that changing the wind and wave conditions separately have on an MABL and the resulting drag relationship between wind and waves, which can also be characterized as the equivalent sea surface roughness. Full-scale MABL domains were set up with $L_x \times L_y \times L_z = 3000\:\text{m} \times 1260\:\text{m} \times 1000\:\text{m}$. Horizontal resolution is $\Delta_x \times \Delta_y = 8\:\text{m} \times 9.8\:\text{m}$; the vertical grid starts with $\Delta_{z,min} = \Delta_x/2 = 4$ m at the bottom of the domain and stretched as described in Section \ref{subsec:validation}. The streamwise resolution was chosen to balance computational expense with resolving a variety of wave spectra without needing to change the resolution, and the vertical resolution at the sea surface was chosen based on a vertical grid convergence study.

Eight MABL cases were run to observe trends based on varying three different parameters: $U_{bulk}$, the bulk wind velocity; $H_s$, the significant wave height of the wave spectrum; and $k_p$, the peak wavenumber of the wave spectrum. A summary of the LES cases run in the parameter sweep is shown in Table \ref{tab:param_sweep}. Conditions used in the horizontal convergence study described in Section \ref{subsec:validation} correspond to the case labeled `Mid'. Wave conditions were chosen based on climatology data from WAVEWATCH III and the JRA55-do reanalysis dataset off the coast of New Jersey \citep{zhou_sea_2023, the_wavewatch_iii_development_group_ww3dg_user_2019, tsujino_jra-55_2018}. A comparison of the LES wind and wave conditions to values of $U_{10}$, $H_s$, and $k_p$ from the climatology data is shown in Figure \ref{fig:climatology}. It demonstrates that, while the MABLs and waves are not in equilibrium because the wave conditions are imposed, none of the combinations of conditions are completely outside the bounds of reality. 

Velocity and vertical momentum flux profiles are shown in Figure \ref{fig:profiles} for cases varying in bulk velocity (the first four cases in Table \ref{tab:param_sweep}). 
Increases in the initial velocity $U_{bulk}$ are, predictably, visible in the mean velocity profile (Figure \ref{fig:profiles}a) but collapse when plotted in viscous units (\ref{fig:profiles}b). The magnitude of the vertical momentum flux also increases with faster bulk velocities (\ref{fig:profiles}c). The decomposition of $xz$ shear stress into subfilter ($\tau_{xz}$) and resolved ($\langle u'w'\rangle$) components is shown in Figure \ref{fig:profiles}d. 

Figure \ref{fig:xyplane} shows examples of instantaneous profiles of surface velocity $\widetilde{u}_b$, filtered wave height $\widetilde{\eta}$, slip parameter $B$ (equation~\eqref{eq:slipB}), dynamic coefficient $\alpha_w$, subfilter roughness $z_{0,\Delta}$, and surface stress (resolved, unresolved, and total) at one spanwise location over the length of the domain's bottom surface for two of the cases in Table \ref{tab:param_sweep}. $\widetilde{\eta}(x,t)$ is prescribed by the defined wave spectrum; $\widetilde{u}_b$ is evaluated at the edge of the first cell, at a height of $\Delta_z = 4$ m above the surface. The right series of panels is from a case with higher $k_p$; the short peak wavelength makes the waves steeper (higher $k_pH_s$). In both columns, there is a qualitative pattern that emerges: local patches with higher velocity also have higher resolved drag, $B$, $\alpha_w$, $z_{0,\Delta}$, and total drag. The case with the steeper waves has generally higher $\alpha_w$ and $z_{0,\Delta}$, as well as generally lower surface velocities. $\alpha_w$ and $z_{0,\Delta}$ vary in space but also between the two cases. The variation of subfilter parameters such as $\alpha_w$ and $z_{0,\Delta}$ in space, unlike variation due to changes with scale, is expected; this variation is neglected in the previous model \citep{aiyer_dynamic_2024}. Although these MABLs are statistically homogeneous, the differences would be more location-dependent in a scenario that involved more heterogeneity.

\section{Momentum Transfer and Surface Stress in LES of Full-Scale MABL}
\label{sec:results}
Section \ref{subsec:ustar} discusses methods for measuring the friction velocity $u_\ast$ in simulations in comparison to in the field. The drag characteristics of the simulated MABLs are discussed and compared to field data collected in the open ocean in Section \ref{subsec:paramsweep} as part of the Risø Air-Sea Experiment \citep[RASEX;][]{mahrt_sea_1996}, Marine Boundary Layer \citep[MBL;][]{hristov_dynamical_2003}, and Coupled Boundary Layers Air-Sea Transfer at Low Winds \citep[CBLAST-LOW;][]{edson_coupled_2007} field experiments, collated and used to develop the Coupled Ocean-Atmosphere Response Experiment (COARE) 3.5 fit \citep{edson_exchange_2013}, as well as data collected during the Gulf of Tehuantepec Experiment \citep[GOTEX;][]{romero_airborne_2010}. The goal with this section is not to achieve a one-to-one match between simulations and field data but to obtain realistic boundary layers in the presence of waves without prescribing the roughness and to understand the sensitivity of the surface drag to changes in wave parameters.

\subsection{Comparing LES to field data}
\label{subsec:ustar}
Measuring the surface stress directly at the ocean surface is not technically feasible, so field measurements calculate the friction velocity based on the vertical momentum flux near the surface assuming that, in the inertial sublayer, \citep{edson_exchange_2013}
\begin{equation}
    \label{eq:us_mom_flux}
    u_\ast^2 \approx \langle u'w'\rangle_s \text{ .}
\end{equation}
Calculating $\langle u'w'\rangle_s$ from a boat or other autonomous platform further requires a constant flux assumption derived from Monin--Obukhov similarity theory: fluxes are constant in the surface layer of a convective atmospheric boundary layer, which is broadly identified as the bottom 10\% of the boundary layer \citep{edson_scalar_2004}. In field experiments such as those cited in \citet{edson_exchange_2013}, boundary layers with a range of stabilities are considered, but the same procedure is used to calculate the friction velocity for all conditions. The MABLs run for the study presented in this paper are all neutral, so the constant stress layer assumption is not valid, as illustrated by the vertical momentum flux profiles in Figure \ref{fig:profiles}, which are linear rather than constant in the bottom 100 m of the MABL. It remains important, however, to keep definitions of quantities such as the friction velocity consistent between simulation and experiment.

\begin{figure}[htbp]
  \centering
  \includegraphics[width=0.7\textwidth]{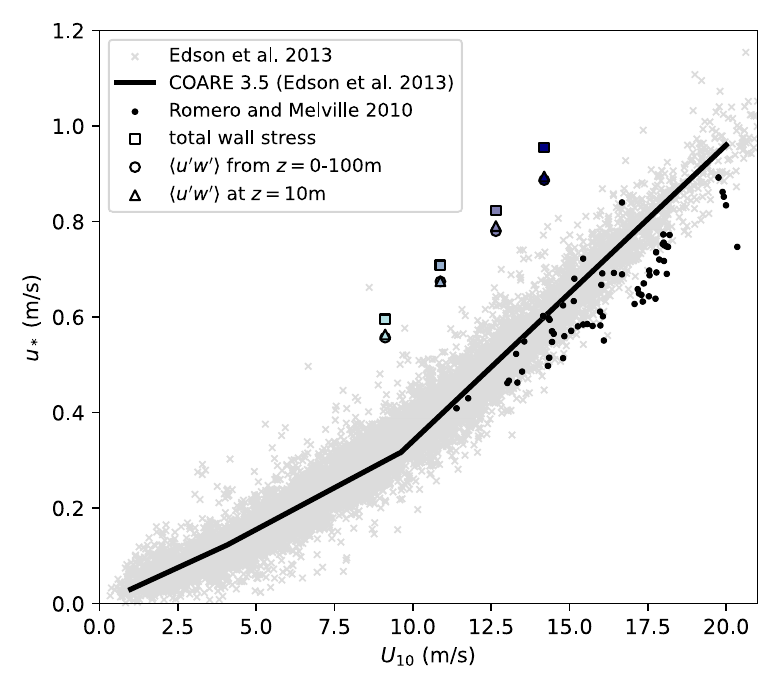}
  \caption{The relationship between the friction velocity $u_\ast$ and wind velocity 10 m above the surface $U_{10}$ for four different wind speeds (bulk velocity $U_{bulk}$ = 15, 18, 21, and 24 m/s from lightest to darkest). The symbols show different methods of calculating $u_\ast$: using the LES total wall stress (squares), $\langle u'w'\rangle$ averaged from 0 to 100m above the sea surface (circles), and $\langle u'w'\rangle$ evaluated at 10 m above the sea surface (triangles). Data from simulations are compared to the COARE 3.5 parameterization \citep{edson_exchange_2013} and data from the GOTEX field campaign \citep{romero_airborne_2010}.}
  \label{fig:ustar}
\end{figure}

In a simulated MABL, the mean (global) friction velocity can be directly determined by probing the total wall shear stress (combined from resolved and subfilter effects) at the surface:
\begin{equation}
    \label{eq:us_les}
  u_\ast^2 = \left\langle \left(\nu + \nu_{sfs}\right)\frac{\partial \widetilde{u}}{\partial z}\bigg|_s \right\rangle + \left\langle \widetilde{F}_d \right\rangle\Delta_z \text{ .}
\end{equation}
Figure \ref{fig:ustar} compares the friction velocity calculated directly from total wall stress (equation~\eqref{eq:us_les}) to friction velocities calculated from the total vertical momentum flux (equation~\eqref{eq:us_mom_flux}, resolved plus subfilter contributions) averaged over the bottom 10\% of the MABL ($z =$ 0-100 m) and calculated at $z=10$ m, as is done in the field experiments. The differences between methods to calculate friction velocity are shown on a plot of $u_\ast$ as a function of $U_{10}$,
which also captures the equivalent global surface roughness (see Figure \ref{fig:z0contours}). The three calculation methods are compared to a commonly-used momentum flux parameterization \citep[COARE 3.5;][]{edson_exchange_2013} as well as representative field data reporting both $u_\ast$ and $U_{10}$ (from the GOTEX experiment \citep{romero_airborne_2010} and from \citet{edson_exchange_2013}). $U_{10}$ is interpolated between grid points in the LES grid. 
Calculating $u_\ast$ through the vertical momentum flux at 10 m and from 0-100 m results in similar estimates, while calculating directly from total wall stress in the LES gives slightly higher drag. Regardless of the method of calculation, the trend with changing velocity $U_{bulk}$ is the same: higher bulk velocity results in movement roughly parallel to the COARE 3.5 line, that is, increases both $U_{10}$ and $u_\ast$ at the rate predicted by the COARE 3.5 fit. 
$u_\ast$ from the LES is generally higher than that estimated by field data for the same $U_{10}$. This could be because of effects not accounted for in the idealized LES such as wave modes that are not aligned with the wind direction or because of a possible underestimation of swell (see Appendix~\ref{app:drag_gap}). We can nonetheless understand how the MABL responds to changing wind and wave conditions from the idealized case. 

\begin{figure}[htbp]
  \centering
  \includegraphics[width=0.8\textwidth]{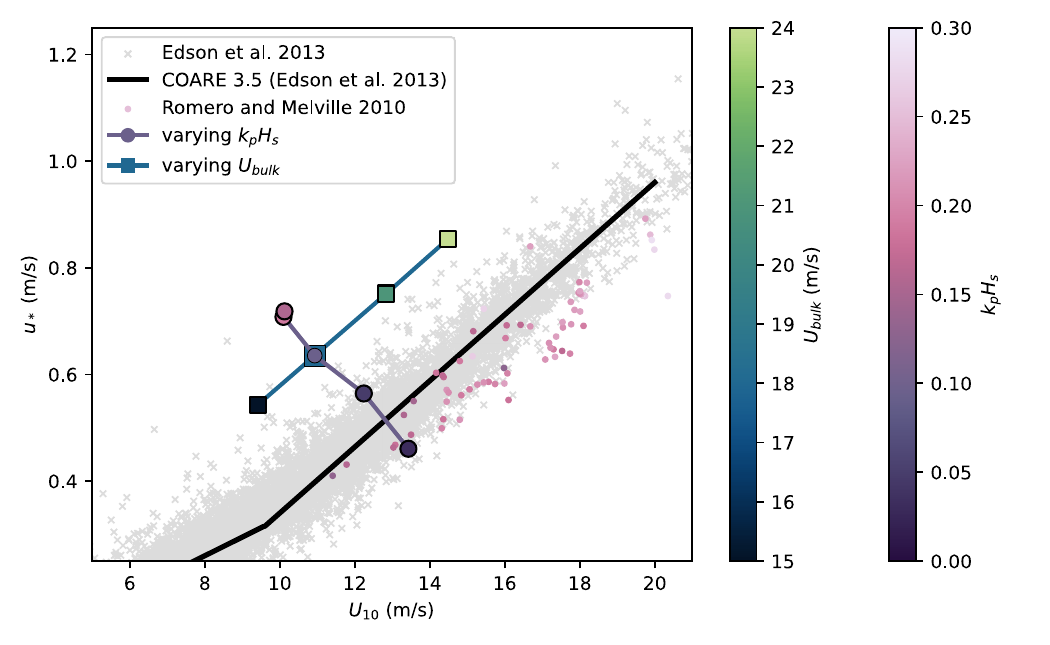}
  \caption{Data from simulations varying wind speed $U_{bulk}$ (blue-green) as well as wave parameters $H_s$ and $k_p$ (purple-pink) compared to the COARE 3.5 fit \citep{edson_exchange_2013} and GOTEX field campaign \citep{romero_airborne_2010}. Variations in wave condition are colored based on the wave steepness $k_pH_s$. The middle case (intersection of the two lines) has nondimensional parameters $H_sk_p$ = 0.094, $c_p/u_\ast = 18.79$, and $c_p/U_{10} = 1.01$.}
  \label{fig:trends}
\end{figure}

\begin{figure}[htbp]
  \centering
  \includegraphics[width=0.8\textwidth]{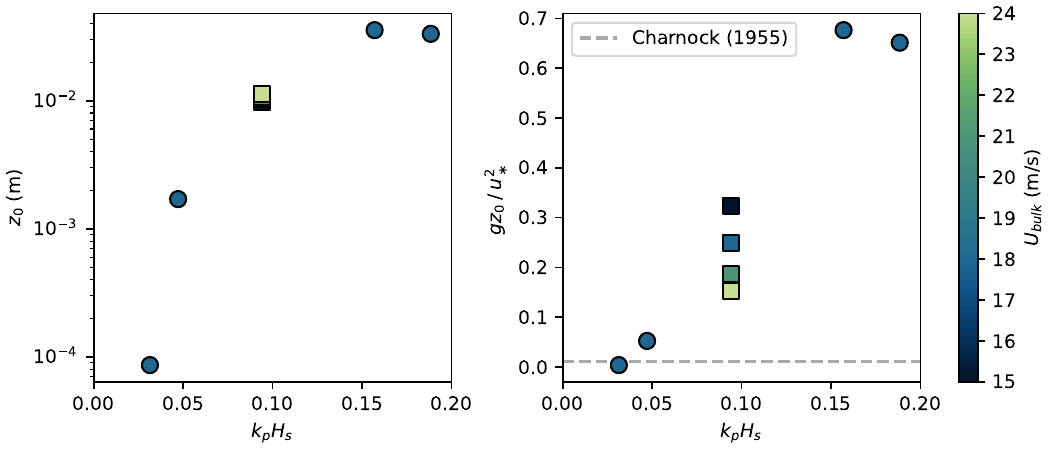}
  \caption{On the left, equivalent global (mean) roughness values assuming a logarithmic profile for cases run in the parameter sweep as a function of $k_pH_s$. In the right plot, the nondimensional value $gz_0/u_\ast^2$, which estimates the value the Charnock parameter would have to be in order to recreate the conditions of the MABL, plotted over $k_pH_s$. The constant estimate from \citet{Charnock1955} for the open ocean is shown with the dashed gray line.}
  \label{fig:swept_z0}
\end{figure}

\subsection{Impact of wind and wave conditions on drag}
\label{subsec:paramsweep}
MABLs generated in the parameter sweep in Table \ref{tab:param_sweep} are compared to field data presented by \citet{edson_exchange_2013} and \citet{romero_airborne_2010} in Figure \ref{fig:trends} in a plot of $u_\ast$ versus $U_{10}$, which characterizes the momentum exchange in a particular wind-wave condition. Values of $u_\ast$ shown in Figure \ref{fig:trends} were computed using the average of $\langle u'w'\rangle$ at 10 m above the surface to match experimental methods; however, the same resulting trends from the parameter sweep analysis were present regardless of $u_\ast$ calculation method. Changes in bulk wind velocity $U_{bulk}$, denoted by blue-green squares, result in variation parallel to the COARE 3.5 fit. However, varying only the wave parameters results in changes across the COARE 3.5 relationship (shown with purple-pink circles). The change in wave parameters can be characterized by the nondimensional parameter $k_pH_s$, a characteristic wave steepness for the spectrum. Increasing $k_pH_s$ involves either increasing the significant wave height $H_s$ for taller waves or increasing the relative velocity between the wind and waves through an increase of the peak wavenumber $k_p$ (a decrease of the peak wavelength $\lambda_p$). It is intuitive that increasing the height of the waves or making the wavelength shorter increases the overall drag: by the dispersion relation for gravity waves, smaller waves travel slower, resulting in a greater velocity difference with the wind.

Changing the wind and wave parameters independently for full-scale MABL simulations without prescribing the friction velocity through a pressure gradient allows for a greater understanding of this surface roughness plot that extends beyond an analysis of the log-law boundary layer relationship, $U_{10} = \frac{u_\ast}{\kappa}\log\left(\frac{10\text{ m}}{z_0}\right)$. Increasing the bulk velocity increases the velocity at 10 m above the surface but simultaneously increases the friction velocity because the waves are slower than the wind by a wider margin. Varying the wind and wave parameters independently from one another reveals that the drag behavior captured by the relationship between $u_\ast$ and $U_{10}$ cannot be captured by a single contour, as is suggested by Charnock and is assumed by many parameterizations of ocean roughness, but instead comprises two distinct impacts of wind speed and characteristic wave steepness. 

Values of an equivalent global $z_0$ as calculated from the logarithmic relationship are plotted in Figure \ref{fig:swept_z0} as a function of $k_pH_s$. Varying the bulk velocity has very little impact on the global roughness, while changing the wave conditions results in global roughness estimates that span several orders of magnitude. \citet{edson_exchange_2013} explores the possibility of roughness dependence on $k_pH_s$ through a linear relationship between the Charnock coefficient and the wave steepness, proposed as $\frac{g z_{0}}{u_\ast} = D k_pH_s$; the computed equivalent Charnock parameter as a function of $k_pH_s$ is plotted on the right panel of Figure \ref{fig:swept_z0} along with the estimate for the open ocean given in \citet{Charnock1955}. The collapse over $U_{bulk}$ that is visible in the left panel no longer occurs because of the Charnock parameter's dependence on $u_\ast$. While it is possible that the relationship is linear, the slope suggested by the LES data is much steeper than that proposed by \citet{edson_exchange_2013}. However, because $u_\ast$ effectively contributes to both axes in the construction of the right panel, the trend is clearer when just the roughness is considered as a function of $k_pH_s$.

In reality, the wind and waves are fully coupled (as opposed to this case, where they are one-way coupled), so the wind and waves are more likely to be in equilibrium with each other; global wave roughness is typically considered dependent on the wind because of this equilibrium.  In the study conducted by \citet{romero_airborne_2010} in particular, the $k_pH_s$ is relatively locked with the wind speed because of the fetch-limited configuration. However, \citet{edson_exchange_2013} find that fully developed seas---defined as those in which the wave age $c_p/u_\ast > 32$---do not occur so commonly in the open ocean because of the ephemerality of storms. \citet{sullivan_large-eddy_2008} also write that disequilibrium is common especially in cases where the wind speed is low or moderate. In the case of this study, the generation of the waves has been completely divorced from that of the wind to separate the primary influence of waves on the MABL from the secondary ones of the completely coupled system. It remains important to be able to separate the distinct influences of wind and waves on boundary layer characteristics in order to make accurate predictions of both local and global momentum transfer between atmosphere and ocean to account for fluctuating winds that may not be in equilibrium with the waves. 

\section{Conclusion}
\label{sec:conclusion}
In this paper, a new formulation of a subfilter wave drag model for LES of wind over ocean waves was developed. The updated local subfilter model includes a slip factor to account for the relative velocity between the wind and waves smaller than the LES filter size. The model was demonstrated to be scale invariant compared to one that neglects the slip. Its locality allows the model to be used to simulate wind conditions in configurations where  
there is some inhomogeneity in the flow: for example an inflow-outflow domain where there is some evolution over the streamwise direction or an LES of an offshore wind farm, where wind and therefore drag conditions in turbine wakes will be very different. 

The updated model was used to run a parameter sweep on full-sized MABLs to compare the relative impacts of varying the wind speed and wave parameters on the drag conditions at the sea surface. The simulated MABLs were compared to field data presented by \citet{edson_exchange_2013} and \citet{romero_airborne_2010}. Comparing the impact of independently changing wind and wave parameters revealed that changes in bulk velocity (wind) caused movement parallel to the relationship defined by the COARE 3.5 fit, while changing the characteristic wave steepness $k_pH_s$ caused movement across the COARE 3.5 line. This implies that, while in a fully-coupled system between the wind and waves there may be some restriction in the space that the surface drag can occupy, the relationship between $u_\ast$ and $U_{10}$ is not one that can be reduced to a simple parameterization and benefits greatly from a local, dynamic implementation of drag that does not require waves to be fully resolved. The foundations of this wave drag model are ready to be implemented for further study of MABL dynamics in offshore wind farms, with changing stability, and more.

\clearpage
\section*{Acknowledgments}
The authors gratefully acknowledge financial support from the Princeton University Andlinger Center for Energy and the Environment, High Meadows Environmental Institute, and the New Jersey Wind Institute Fellowship Program. This work is supported by the National Science Foundation under grant 2318816 to LD (Physical Oceanography program).

%
%
\section*{Data Statement}
The data that support the findings of this study are available at \url{https://github.com/hhwilliams/slip-subgrid-wave-drag}. 

\clearpage
\section*{Appendices}
\addcontentsline{toc}{section}{Appendices}
\setcounter{subsection}{0}
\renewcommand{\thesubsection}{\Alph{subsection}}
\titleformat{\subsection}[block]{\normalfont\large\bfseries}{Appendix \thesubsection:}{0.5em}{}
\counterwithin*{figure}{subsection}
\counterwithin*{table}{subsection}
\renewcommand{\thefigure}{\thesubsection\arabic{figure}}
\renewcommand{\thetable}{\thesubsection\arabic{table}}
\renewcommand{\theHfigure}{app.\thefigure}
\renewcommand{\theHtable}{app.\thetable}
\renewcommand{\theHsubsection}{app.\thesubsection}
\subsection{Drag gap between wind waves and swell waves for resolved wave drag model}
\label{app:drag_gap}
\begin{figure}[htbp]
  \centering
  \includegraphics[width=0.9\textwidth]{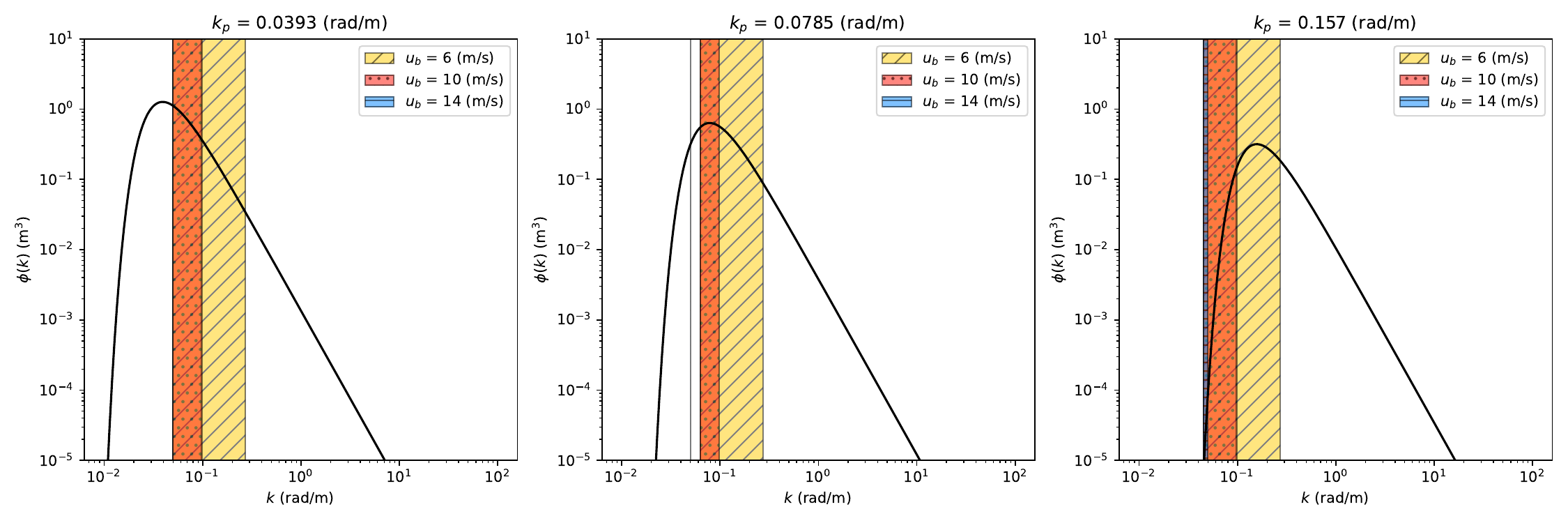}
  \caption{
  Ranges of wave modes where there is no drag shaded for $k_p$ = 0.0393 rad/m (left), 0.0785 rad/m (center), and 0.157 rad/m (right) given three different surface velocities $u_b$, denoted by color and hatching so overlap is visible. Example wind velocities are taken from the approximate range seen in the simulation with $U_{bulk} = 18$ m/s; approximate global $u_\ast$ from the LES for each wave condition run for the parameter study was used to determine swell limits. }
  \label{fig:nodrag_zone}
\end{figure}

The wave drag model for resolved waves presented in equation~\eqref{eq:res_waves} includes an intermediate wavenumber region between wind wave and swell regimes where there is no drag imparted on the MABL (swell waves impart ``negative drag'', or accelerate the flow). These wave modes are not traveling slower than the surface wind speed but are also not traveling fast enough such that they satisfy the condition put on swell waves, that is, $c/u_\ast > 25$ \citep{aiyer_dynamic_2024}. The high-wavenumber boundary of the band without drag is determined by the surface velocity of the flow and the condition that $u_b > c_n$ for wind waves (slower than the wind). The low-wavenumber boundary is determined by the condition that $c/u_\ast > 25$ and an additional condition that $k < 2k_p$ to prevent against very small and slow waves injecting momentum; in practice, the first condition is more restrictive than the second so limits the range of wave modes injecting momentum. Momentum is not exchanged the same way in the positive and negative directions due to the asymmetry of waves, and the resolved wave drag model is not intended to model the acceleration of a quiescent boundary layer. Examples of wavenumber ranges where there is no drag calculated by the resolved drag model \citep{aiyer_dynamic_2024} are shown in Figure \ref{fig:nodrag_zone}, separated by wave condition and for several surface wind velocities. Global $u_\ast$ values are used from the LES runs with $U_{bulk}$ = 18 m/s and the according wave condition. It is possible that the condition on swell modes artificially restricts momentum injection into the wind from modes traveling faster than the surface velocity, potentially contributing to a systematic overestimation of drag due to the missing negative drag of these intermediate waves. This could explain the overestimation of $u_\ast$ in Figures \ref{fig:ustar} and \ref{fig:trends}. 

\subsection{Behavior and form of weighted subfilter wavenumber \texorpdfstring{$k_{sf}$}{k\_sf} and slip model parameters}
\label{app:ksf}
\begin{figure}[htbp]
  \centering
  \includegraphics[width=0.5\textwidth]{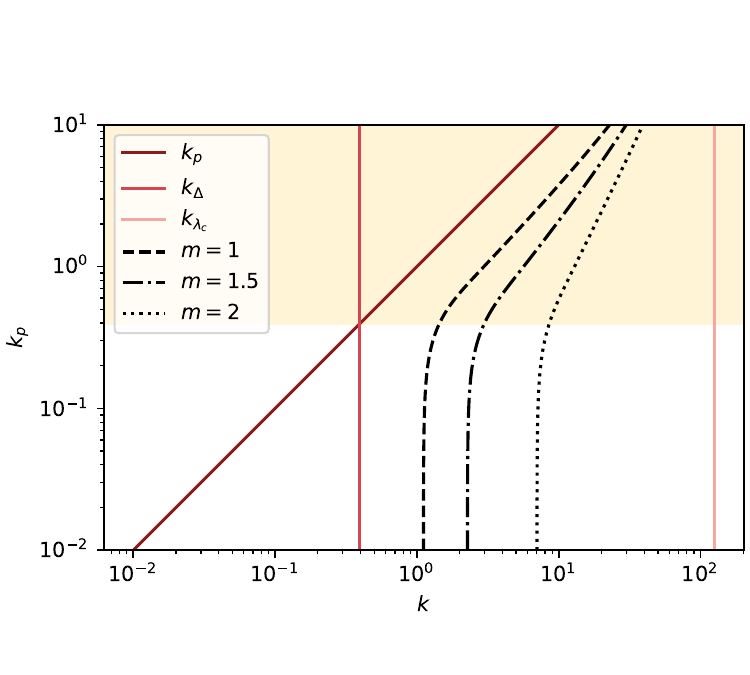}
  \caption{
  Behavior of $k_{sf}$ with changing $k_p$ for different $k_{sf}$ model constructions. The shaded region represents under-resolved scenarios ($k_p > k_\Delta$) where the model will still converge but not produce accurate results. }
  \label{fig:sfk_sandbox}
\end{figure}

\begin{figure}[htbp]
  \centering
  \includegraphics[width=0.9\textwidth]{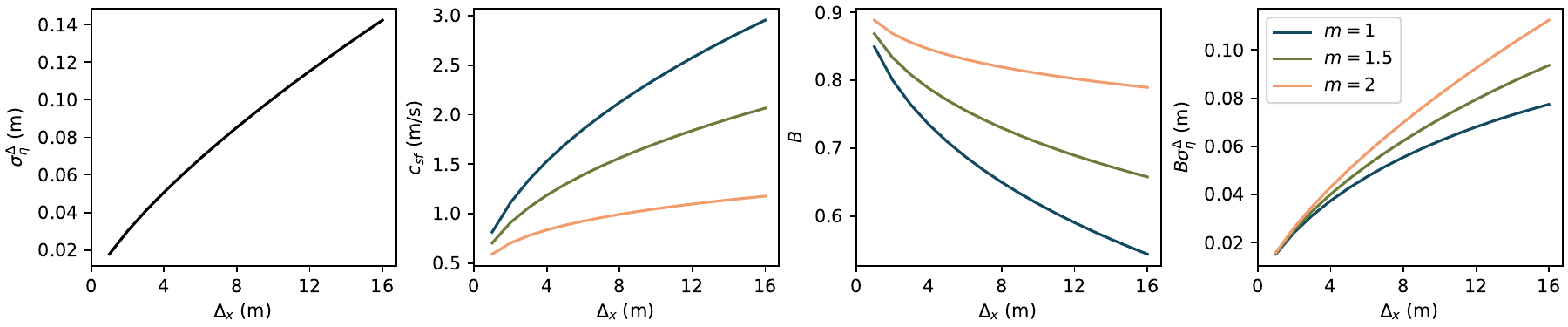}
  \caption{
  From left to right: behavior of $c_{sf}$, $B$, $\sigma_\eta^\Delta$, and $B \sigma_\eta^\Delta$ as a function of $\Delta_x$ for a given $\widetilde{u}_b$, $k_p$, and $H_s$. }
  \label{fig:sf_model_behavior}
\end{figure}

Early investigations included $k_{sf}$ equal to $k_\Delta$ as well as the weighted formulations of $k_{sf}$ mentioned in Section \ref{subsec:slip}. Using $k_\Delta$ as the characteristic subfilter wavenumber resulted in dynamic procedures that frequently did not converge.  Physically, it is intuitive that a representative subfilter $k_{sf}$ would exist somewhere between $k_\Delta$ and $k_{\lambda_c}$. Figure \ref{fig:sfk_sandbox} visualizes how differently-weighted $k_{sf}$ (that is, $k_{sf}$ with different values of $m$) shift the characteristic subfilter wavenumber between $k_\Delta$ and $k_{\lambda_c}$; as $m$ increases, $k_{sf}$ moves closer to $k_{\lambda_c}$ and the representative subfilter wave mode becomes smaller, closer to the capillary length scale.

Figure \ref{fig:sf_model_behavior} shows how the behavior of other factors in the subfilter wave drag model with slip change with changing $\Delta_x$ for each of the weighted $k_{sf}$ models.  These were all computed with constant $\widetilde{u}_b$ = 10 m/s, $k_p$ = 0.0785 rad/m, and $H_s$ = 1.2 m. The amount of variation in $B$ is controlled by $c_{sf}$; the variation in $B\sigma_\eta^\Delta$ is weakest with $m = 1$ because the variation in $B$ and $\sigma_\eta^\Delta$ counter each other.

\subsection{Comparison of local, slip subfilter wave drag model performance to global, no slip model}
\label{app:aiyer_validation}

\begin{figure}[htbp]
    \centering
    \includegraphics[width=0.9\linewidth]{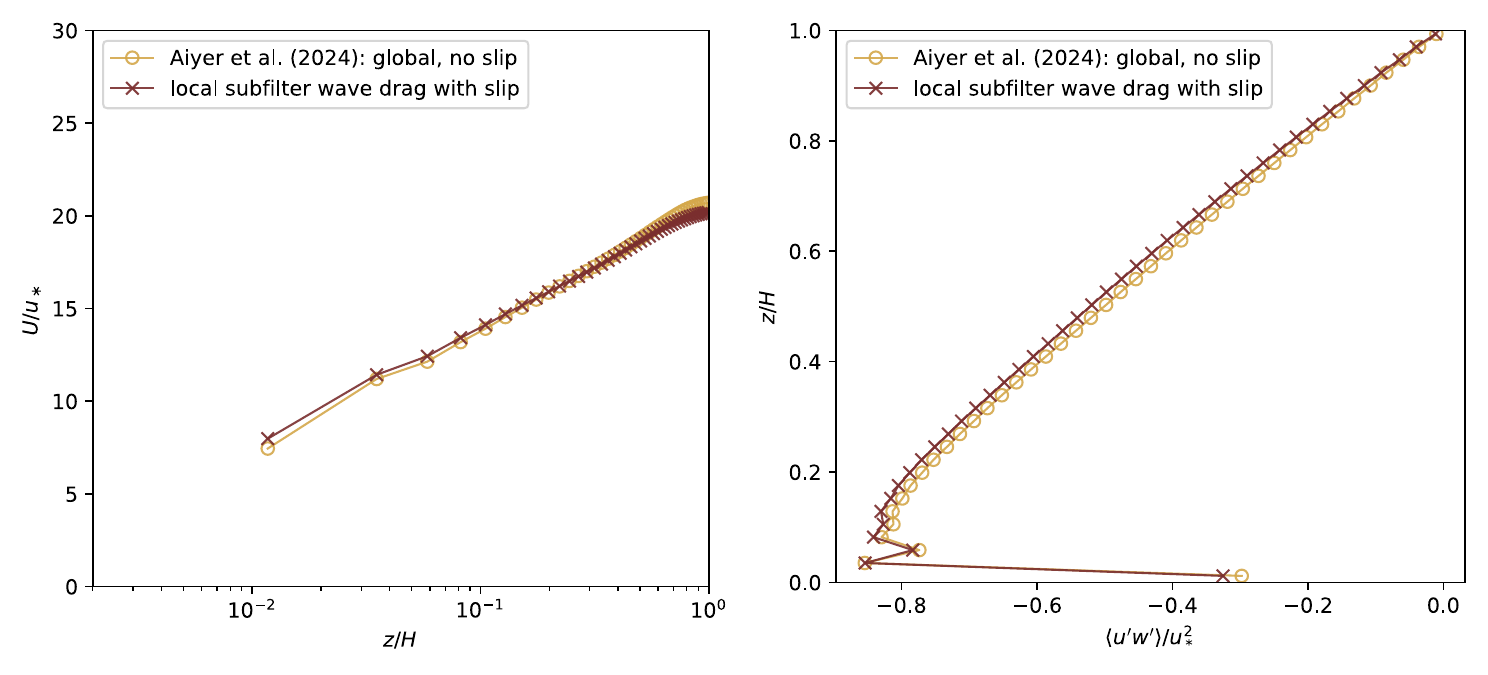}
    \caption{
    Comparison of global no slip model presented in \citet{aiyer_dynamic_2024} to local slip model in this paper for case run with constant pressure gradient such that $u_\ast$ = 0.443 m/s and JONSWAP wave spectrum with $U_{10}$ = 12 m/s and fetch $X$ = 7505.5 m following \citet{aiyer_dynamic_2024}. Mean velocity is shown on the left and vertical momentum flux on the right. 
    }
    \label{fig:validation}
\end{figure}

To ensure that the local slip model performs as expected in known scenarios, one of the cases presented in \citet{aiyer_dynamic_2024}---run with a global, no slip model--- was recreated. This case was driven with a constant pressure gradient, which sets the friction velocity through a force balance such that 
\begin{equation}
    \frac{\partial P}{\partial x} = -\frac{u_\ast^2\rho}{H} \text{ ,}
    \label{eq:pgrad}
\end{equation}
where $P$ is the pressure, $\rho_a$ is the air density, and $H$ is the boundary layer height. The pressure gradient is set such that $u_\ast$ = 0.443 m/s. The one-dimensional wave field is set according to the JONSWAP spectrum \citep{hasselmann_measurements_1973} with $U_{10}$ = 12 m/s and $X$ = 7505.5 m; these conditions ensure that the wave age $c_p/u_\ast$ = 10. The domain has dimensions $L_x \times L_y \times L_z = 62.8\text{ m} \times 62.8 \text{ m} \times 25.1 \text{ m}$. More details such as vertical grid stretching and velocity initialization follow the case labeled CU10(C) in \citet{aiyer_dynamic_2024} and can be found there. 

The mean velocity profile (left) and vertical momentum flux (right) from running the same scenario with the global, no slip subfilter model \citep{aiyer_dynamic_2024} and the updated local slip subfilter model are shown in Figure \ref{fig:validation}. There is good agreement in both the mean velocity and resolved vertical momentum flux between the two models in these heterogeneous conditions. 

\subsection{Pressure Gradient-Driven MABLs}
\label{app:pgrad}
\begin{figure}[htbp]
  \centering
  \includegraphics[width=0.95\textwidth]{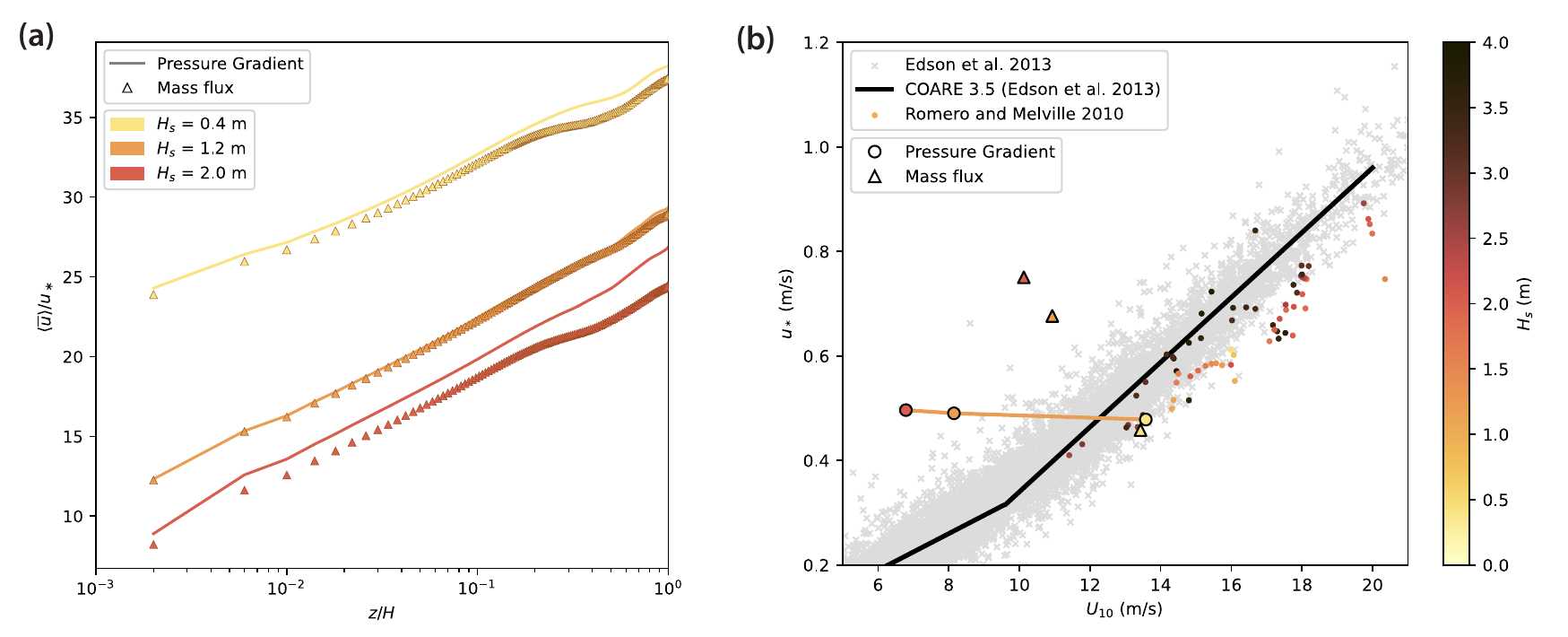}
  \caption{Mean velocity profiles from for pressure gradient-driven LES cases (lines) compared to `Mid' case in Table \ref{tab:param_sweep}, driven with constant mass flux (triangles). The two orange cases have the same wave conditions ($H_s$ = 1.2 m, $k_p$ = 0.0785 rad/m) (a). Friction velocity as a function of $U_{10}$ for pressure gradient-driven LES cases plotted with experimental data from \citet{edson_exchange_2013} and \citet{romero_airborne_2010} as well as `Mid' case from Table \ref{tab:param_sweep} (b).}
  \label{fig:pgrad}
\end{figure}

Previous LES studies on the air-sea interface have generally used an imposed pressure gradient to drive the flow, constraining the global friction velocity through Equation~\eqref{eq:pgrad}. Prescribing global $u_\ast$ in this way predetermines an aspect of the drag between the wind and waves; doing so has resulted in learning about the nature of the form drag on and turbulent motions above the waves but makes it difficult to study how the shape of the MABL changes with wave conditions since this is largely a function of the total stress. 

Cases varying the significant wave height $H_s$ were also run with an imposed pressure gradient such that $u_\ast$ = 0.5 m/s. Figure \ref{fig:pgrad}a shows the mean velocity profiles from these cases alongside the associated constant mass flux cases (those with the same wave conditions), colored by $H_s$. Mean velocity profiles for MABLs run with constant mass flux are very similar to those run with a pressure gradient except at the very top of the domain, which is very far from the surface. The profiles are not expected to line up perfectly in this plot since the pressure gradient cases have predefined $u_\ast$.

Figure \ref{fig:pgrad}b shows the relationship between $U_{10}$ and $u_\ast$ for the pressure gradient-driven cases alongside the COARE 3.5 fit and field data from \citet{edson_exchange_2013} and \citet{romero_airborne_2010}. Examining the drag trend through the pressure gradient-driven cases tells a somewhat different story than that presented with the constant mass flux cases in Section \ref{sec:results} because $u_\ast$ is constrained. The trend in wind velocity $U_{10}$ based on changing $H_s$ is in the same general direction as Figure \ref{fig:trends}: larger waves slow the wind since they contribute more to the fixed drag. The movement is only constrained to the one axis in this plot, however, and is seen only in a change in wind velocity because the global friction velocity is fixed. The change in $U_{10}$ spans 7 m/s rather than the 4 m/s spanned by changing wave parameters in the constant mass flux LES cases shown in Figure \ref{fig:trends}. 
Interestingly, the equivalent global surface roughness, shown in Table \ref{tab:pgrad}, remains similar between the methods of driving the flow. 

Constraining the simulation through the predetermination of a friction velocity masks the physical effect of the waves as an influence only on the velocity rather than also a physical change to the total drag. Because the analysis of interest is in the surface roughness and shape of the MABL profile near the surface, using simulations driven with constant mass flux (constant bulk velocity) rather than an imposed pressure gradient provide a more complete physical picture. 

\begin{table}[htbp]
    \caption{Equivalent global roughness $z_0$, in meters, for cases driven with pressure gradient compared to those run with constant mass flux. }
  \begin{center}
    \begin{tabular}{ c | *{3}{c}}
      \hline 
       & $H_s$ = 0.4 m & $H_s$ = 1.2 m & $H_s$ = 2.0 m \\
       \hline
       Pressure gradient & $1.17\times 10^{-4}$ & $1.30\times 10^{-2}$ & $4.21\times 10^{-2}$ \\
       Mass flux & $8.12\times 10^{-5}$ & $1.56\times 10^{-2}$ & $4.53\times 10^{-2}$ \\
       \hline
    \end{tabular}
  \end{center}
  \label{tab:pgrad}
\end{table}

\clearpage
\bibliography{references}

\end{document}